\documentclass[final,3p,times]{elsarticle}
\usepackage{amsfonts}
\usepackage{mathrsfs}
\usepackage{amsmath}
\usepackage{xcolor}
\usepackage{amssymb}
\usepackage{bm}
\usepackage{feynmp}
\usepackage{extarrows}
\usepackage{slashed}
\usepackage{graphicx}
\usepackage{subfigure}
\usepackage{verbatim} 
\usepackage{ulem} 
\usepackage{float}
\usepackage[linktocpage=true,bookmarks=true,bookmarksnumbered=true,breaklinks=true,pdfpagemode=Fullscreen,pdfstartview=FitBH]{hyperref}

\renewcommand{\rm}{\mathrm}
\def\bge{\begin{equation}}
\def\ede{\end{equation}}
\def\bga{\begin{aligned}}
\def\eda{\end{aligned}}
\newcommand{\beq}{\begin{equation}}
\newcommand{\eeq}{\end{equation}}
\newcommand{\bq}{\begin{equation}}
\newcommand{\eq}{\end{equation}}
\newcommand{\ba}{\begin{array}}
\newcommand{\ea}{\end{array}}
\newcommand{\beqa}{\begin{eqnarray}}
\newcommand{\eeqa}{\end{eqnarray}}
\newcommand{\beqs}{\begin{subequations}}
\newcommand{\eeqs}{\end{subequations}}

\def\nn{\nonumber}

\def\({\left(}
\def\){\right)}
\def\[{\left[}
\def\]{\right]}

\def\End{\end{document}}

\def\leqq{\leqslant}
\def\geqq{\geqslant}

\def\al{\alpha}

\def\ga{\gamma}

\def\tanb{\tan\!\beta}
\def\tanB{\tan\!\beta}
\def\cosba{\cos(\beta\!-\!\alpha)}

\def\sla{\!\!\!\slash}

\def\to{\rightarrow}

\def\gaga{\gamma\gamma}

\newcommand{\mL}{\mathcal{L}}
\newcommand{\mO}{\mathcal{O}}

\newcommand{\TeV}{\textrm{TeV}}
\newcommand{\GeV}{\textrm{GeV}}
\def\gaga{\gamma\gamma}
\def\bb{b\bar{b}}

\def\Ha{\mathbb{H}_1^{}}
\def\Hb{\mathbb{H}_2^{}}
\def\Had{\mathbb{H}_1^{\dag}}
\def\Hbd{\mathbb{H}_2^{\dag}}
\def\ZZ{\mathbb{Z}}
\def\tanb{\tan\beta}

\def\End{\end{document}}

\begin{document}

\begin{frontmatter}

\setcounter{footnote}{1}
\renewcommand{\thefootnote}{\fnsymbol{footnote}}

\title{{\bf Searching Heavier Higgs Boson via Di-Higgs Production at LHC Run-2}}

\author{
{\sc Lan-Chun L\"{u}}\,\footnote{ lvlc10@mails.tsinghua.edu.cn}$^{a}$,~~
{\sc Chun Du}\,\footnote{ chun.thazen.du@gmail.com}$^{b}$,~~
{\sc Yaquan Fang}\,\footnote{fangyq@ihep.ac.cn (corresponding author)}$^{b}$,~~
{\sc Hong-Jian He}\,\footnote{hjhe@tsinghua.edu.cn (corresponding author)}$^{a,c,d}$,~~
{\sc Huijun Zhang}\,\footnote{huijun.zhang@cern.ch}$^{e}$
\vspace*{3mm}
}

\address{
$^a$\,Institute of Modern Physics and Center for High Energy Physics,
Tsinghua University, Beijing 100084, China
\\[1mm]
$^b$\,Institute of High Energy Physics, Beijing 100049, China
\\[1mm]
$^c$\,Institute for Advanced Study, Princeton, New Jersey 08540, USA
\\[1mm]
$^d$\,Harvard University, 1 Oxford Street, Cambridge, Massachusetts 02138, USA
\\[1mm]
$^e$\,Nanjing University, Nanjing 210093, China
}

\begin{abstract}
The discovery of a light Higgs particle $\,h^0$\,(125\,GeV)\, opens up new prospect for searching heavier
Higgs boson(s) at the LHC Run-2, which will unambiguously point to new physics beyond the standard model (SM).
We study the detection of a heavier neutral Higgs boson $\,H^0$ via di-Higgs production channel
at the LHC\,(14\,TeV), $\,H^0 \!\!\to\! h^0h^0 \!\!\to\! WW^*\gamma\gamma$\,.\,
This directly probes the $\,Hhh\,$ cubic Higgs interaction, which exists in most extensions
of the SM Higgs sector. For the decay products of final states $\,WW^*$,\,
we include both pure leptonic mode $\,WW^* \!\!\to\! \ell\bar{\nu}\bar{\ell}\nu\,$
and semi-leptonic mode $\,WW^* \!\!\to\! q\bar{q}'\ell\nu$\,.\,
We analyze signals and backgrounds by performing fast detector simulation for the full process
$\,pp \!\to\! H \!\to\! hh \!\to\! WW^*\gamma\gamma \!\to\! \ell\bar{\nu}\bar{\ell}\nu\gamma\gamma\,$
and
$\,pp \!\to\! H \!\to\! hh \!\to\! WW^*\gamma\gamma \!\to\! \ell\nu q\bar{q}'\gamma\gamma$,\,
over the mass range $M_H^{}=250-600$\,GeV.\,
For generic two-Higgs-doublet models (2HDM), we present the discovery reach of
the heavier Higgs boson at the LHC Run-2, and compare it with
the current Higgs global fit of the 2HDM parameter space.
\\[1.5mm]
\\[1.5mm] PACS numbers: 12.15.Ji, 12.60.-i, 12.60.Fr, 14.80.Ec
\hfill Physics Letters B (2016)~$[$\,arXiv:1507.02644\,$]$
\end{abstract}


\end{frontmatter}



\renewcommand{\thefootnote}{\arabic{footnote}}

\section{Introduction}

Most extensions of the standard model (SM) require an enlarged Higgs sector, containing more than
one neutral Higgs states. After the LHC discovery of a light Higgs particle $\,h^0$\,(125\,GeV)\,
\cite{ATLAS2012}\cite{CMS2012}, a pressing task of the on-going LHC Run-2
is to search for additional new Higgs boson(s),
which can unambiguously point to new physics beyond the SM.

\vspace*{1mm}

Such an enlarged Higgs sector\,\cite{exHiggs}
may contain additional Higgs doublet(s), or Higgs triplet(s), or Higgs singlet(s).
For instance, the minimal supersymmetric SM (MSSM)\,\cite{MSSM} always requires two Higgs doublets
and its next-to-minimal extension (NMSSM)\,\cite{NMSSM} further adds a Higgs singlet.
The minimal gauge extensions with extra SU(2) or U(1) gauge group\,\cite{SU2x}\cite{U1x}
will invoke an additional Higgs doublet or singlet.
The minimal left-right symmetric models\,\cite{LR} include an extra product group
$\text{SU(2)}_R^{}\otimes \text{U(1)}_{B-L}^{}$,
and thus requires a Higgs bidoublet plus two Higgs triplets.
For the demonstration in our present LHC study,
we will consider generic two-Higgs-doublet models (2HDM) \cite{Branco:2011iw} under the SM gauge group.
To evade constraints of flavor changing neutral current (FCNC),
it is common to impose a discrete $\ZZ_{2}^{}$
symmetry on the 2HDM. For different model settings of Higgs Yukawa interactions,
the 2HDMs are conventionally classified into type-I, type-II, lepton-specific, neutrino-specific,
and flipped 2HDMs \cite{Branco:2011iw}.
The current study will focus on the conventional type-I and type-II 2HDMs.

\vspace*{1mm}

For the heavier Higgs state $H^0$ with mass above twice of the light Higgs boson $h^0$,
$\,M_H^{}> 2M_h^{}\simeq 250\,$GeV,\, the di-Higgs decay channel $\,H\to hh\,$ is opened and
becomes significant, in addition to
the other SM-like major decay modes $\,H\to WW, ZZ\,$.\,
Hence, the LHC can search for the di-Higgs production channel
$\,pp\to H\to hh\,$ \cite{MSSM,SU2x,Wells07,hhh-bbgg}.
ATLAS analyzed the decay channel $\,hh\to \bb\gamma\gamma\,$ at the LHC\,(8\,TeV) run and found
a $2.4\sigma$ excess at $\,M(\bb\gaga)\simeq 300$\,GeV \cite{Aad:2014yja}.
CMS performed similar searches
for this channel and derived limits on the parameter space \cite{CMS:2013eua}.
An analysis of this channel at 14\,TeV runs with high luminosity 1000\,fb$^{-1}$
was done for 2HDM \cite{bbgaga}.
Another study considered the SM plus a heavy singlet scalar via
$\,H\to hh\to \bb WW^*\to \bb \ell\nu\ell\nu\,$  channel
for 14\,TeV runs with $3000\mathrm{fb}^{-1}$ luminosity \cite{bbWW}.
We note that it is possible to increase the sensitivity of $H^0$ searches by
studying and combining more decay channels of the di-Higgs bosons.

In this work, we perform systematical study of $\,H^0\,$ production
via a new decay channel of di-Higgs bosons,
$\,pp\!\to\! H\!\to\! hh\!\to\! WW^*\gamma\gamma\,$.\,
For the final state weak bosons, we will analyze both
pure leptonic mode $\,WW^*\!\to\ell\bar{\nu}\bar{\ell}\nu\,$ and
semi-leptonic mode $\,WW^*\!\to q\bar{q}'\ell\nu\,$.\,
Since a SM-like Higgs boson $\,h^0$(125GeV)\, has decay branching fractions
$\,\text{Br}[h\to \bb , WW^*\!,\,ZZ^*]\simeq (58\%,\,22.5\%,\,2.77\%)$,\,
we see that the di-Higgs decay mode $\,hh\rightarrow WW^*\gamma\gamma\,$
(with pure leptonic or semi-leptonic $WW^*$ decays) has
the advantage of much cleaner backgrounds than $\,hh\rightarrow \bb\gamma\gamma\,$,\,
while $\,\text{Br}[h\to WW^*]\,$ is only smaller than $\,\text{Br}[h\to \bb ]\,$
by about a factor of $\,2.6\,$.\,
Hence, we expect that the $\,hh\rightarrow WW^*\gamma\gamma\,$ mode should have
comparable sensitivity to $\,hh\rightarrow \bb\gamma\gamma\,$ mode, and is more sensitive
than $\,hh\rightarrow \bb WW^*\,$ mode.

\vspace*{1mm}

This letter paper is organized as follows. In section\,\ref{sec:2},
we present the production and decays
of the heavier Higgs boson $H^0$ in 2HDM type-I and type-II.
Then, in section\,\ref{sec:3},
we systematically analyze the signals and backgrounds for the reaction
$\,pp\to H \to hh\rightarrow WW^*\gamma\gamma\,$,\,
including both pure leptonic and semi-leptonic decay modes of the $WW^*$ final state.
In section\,\ref{sec:4}, we further analyze the LHC probe of the parameter space
for 2HDM-I and 2HDM-II, and compare it with the current Higgs global fit.
Finally, we conclude in section\,\ref{sec:5}.

\vspace*{2mm}
\section{Decays and Production of Heavier Higgs Boson $\,H^0\,$ in the 2HDM}
\label{sec:2}

\vspace*{2mm}
\subsection{{\bf 2HDM Setup and Parameter Space}}
\vspace*{2mm}

For the present phenomenological study,
we consider the 2HDM \cite{Branco:2011iw} as the minimal extension of the SM Higgs sector.
We set the Higgs potential to have CP conservation, and the two Higgs doublets $\,\Ha\,$ and $\,\Hb\,$ have
hypercharge $\,Y=+\frac{1}{2}\,$,\, under the convention $\,Q=I_3^{}+Y\,$.\,
It is desirable to assign a discrete $\,\ZZ_{2}^{}$\, symmetry to the Higgs sector,
under which the Higgs doublet  $\,\Ha\,(\,\Hb)\,$  is $\,\ZZ_{2}^{}$ even (odd).
With these, the Higgs potential can be written as
\beqa
\label{eq:2HDM_potential}
V &=& M_{11}^2|\Ha|^2 + M_{22}^2|\Hb|^2 - M_{12}^2(\,\Had\Hb +\Hbd\Ha)
+\frac{\lambda_1^{}}{2}(\Had\Ha)^2 + \frac{\lambda_2^{}}{2}(\Hbd \Hb)^2
\nn \\
&& + \lambda_3^{}\,|\Ha|^2 |\Hb|^2+\lambda_4^{} |\Had \Hb|^2
   + \frac{\lambda_{5}^{}}{2}\Big[(\Had\Hb )^2+ (\Hbd\Ha )^2\Big]\,,
\eeqa
where the masses and couplings are real,
and we have allowed a soft $\,\ZZ_{2}^{}\,$ breaking mass term of
$\,M_{12}^2\,$.\,  The minimization of this Higgs potential gives
the vacuum expectation values (VEVs),
$\,\langle\Ha\rangle = \frac{1}{\sqrt{2}\,}(0,\,v_1^{})^T\,$ and
$\,\langle\Hb\rangle = \frac{1}{\sqrt{2}\,}(0,\,v_2^{})^T$.\,
The two doublets jointly generate the electroweak symmetry breaking (EWSB) VEV $\,v\simeq 246\,$GeV,
via the relation  $\,v=(v_1^2+v_2^2)^{1/2}_{}\,$,\,
where $\,v_{1}^{} =v\cos\beta$\, and $\,v_{2}^{} =v\sin\beta$\,.\,
Thus, the parameter $\,\tan\beta\,$ is determined by the Higgs VEV ratio,
$\,\tan\beta = {v_2^{}}/{v_1^{}}\,$.\,
The two Higgs doublets contain eight real components in total,
\begin{eqnarray}
\mathbb{H}_{j}^{}  \,=\, \left(\!\! \begin{array}{c} \pi_{j}^{+}
\\ \frac{1}{\sqrt{2}\,} \(v_{j}^{} + h_j^{} +i\pi_j^{0}\) \\
\end{array}  \!\!\right) , ~~~~~~~(j=1,2)\,.
\end{eqnarray}
Three imaginary components are absorbed by $(W^{\pm},\,Z^0)$ gauge bosons,
while the remaining five components give rise to the two CP-even neutral states $(h^0_1,\,h_2^0)$,
one CP-odd neutral state $A^0$, and two charged states $H^{\pm}$.\,
The mass eigenstates $(h,\,H)$ of the neutral CP-even Higgs bosons are given by
diagonalizing the mass terms in the Higgs potential (\ref{eq:2HDM_potential}).
They are mixtures of the gauge eigenstates $\,(h_{1}^{},\,h_{2}^{})$\,,
\beqa
\label{eq:evenHiggs}
\left(\!\!\begin{array}{c}
h \\[1mm] H   \end{array} \!\right)
~=~
\left(\!\!\begin{array}{rr}
\cos\al & -\sin\al \\[1mm]
\sin\al & \cos\al
\end{array} \!\right)
\left(\!\begin{array}{c} h_2^{} \\[1mm]
h_1^{} \end{array} \!\!\right)\, ,
\eeqa
where 
$\,\alpha\,$ is the mixing angle.
Among the two neutral Higgs bosons,
$\,h\,$ is the SM-like Higgs boson with mass $\,M_h^{}\simeq 125\,\GeV$,\,
as discovered at the LHC Run-1 \cite{ATLAS2012}\cite{CMS2012}, and $\,H\,$ is the heavier Higgs state.
We will systematically study the LHC discovery potential of $\,H\,$ state in the present work.
The Higgs potential \eqref{eq:2HDM_potential} contains $8$ parameters in total,
three masses and five couplings.
Among these, we redefine 7 parameters as follows:
the EWSB VEV $v$,\, the VEV ratio $\,\tanB\,$,\,
the mixing angle $\,\alpha$,\, and the mass-eigenvalues
\,$(M_h^{}, M_H^{},\, M_A^{},\, M_{H\pm}^{})$.\,
We may choose the 8th parameter as the Higgs mass-parameter $\,M_{12}^2\,$.\,
Note that once we fix the mass spectrum of the 5 Higgs bosons as inputs,
we are left with only 3 independent parameters $\,(\alpha \, ,\tanB )\,$  and $\,M_{12}^2\,$.\,
The current LHC data favor the parameter space of the 2HDM around the alignment limit \cite{Branco:2011iw},
under which $\,\cos (\beta -\alpha) =0\,$.\, This limit corresponds to the light Higgs state $\,h\,$
to behave as the SM Higgs boson with mass 125\,GeV.\,
For practical analysis, we fix $\,M_h^{}\simeq 125\,$GeV\, by the LHC data and vary the heavier
mass $M_H^{}$ within the range of $\,250-600$\,GeV.\,
We consider the Higgs states $A$ and $H^\pm$ to be
relatively heavy, within the mass-range $\,M_A^{}, M_{H\pm}^{}=0.3-2\,$TeV for simplicity.
We will scan the parameter space and analyze the LHC production and decays of $\,H\,$ in the
next section.

The heavier neutral Higgs boson \,$H$\, has gauge couplings with $(W^{\pm},\,Z^0)$ and
Yukawa couplings with quarks and leptons, which depend on the VEV ratio
$\,\tanB\,$ and mixing angle $\,\alpha\,$.\,
The gauge couplings of $\,H\,$ with $\,V\,(=W,Z)$ differ from the SM Higgs coupling
by a scaling factor $\,\cosba\,$,
\beqa
\label{eq:H_gauge}
G_{HVV}^{} \,=\, \cosba \,G_{HVV}^{\text{sm}}\,,\qquad~~~
G_{HVV}^{\text{sm}} \,=\, \frac{\,2 M_{V}^{2}\,}{v} \,.
\eeqa
The Yukawa interactions of $\,H\,$ with fermions can be expressed as follows,
\beqa
\label{eq:H_Yukawa}
\mL_{\text{Y}(H)}^{} \,=~ -\! \sum_{f=u,d,\ell}\frac{\,m_f^{}\,}{v}\xi_{H}^{f}\, \bar f f\,H \,,
\eeqa
where the dimensionless coefficient $\,\xi_{H}^{f}\,$
differs between the Type-I and Type-II of 2HDM, as summarized in Table\,\ref{tab:1}.

\begin{table}[h]
\centering
\caption{Summary of the Yukawa couplings $\,\xi_H^f\,$ between the heavier Higgs boson $H^0$ and
the SM fermions in 2HDM-I and 2HDM-II, where we have factorized out a common factor $\,m_f^{}/v\,$
(corresponding to the SM Higgs Yukawa coupling).}
\vspace*{1.5mm}
\begin{tabular}{c||c|c|c}
\hline\hline
&&&\\[-2.5mm]
 2HDM &  $\xi_{H}^{u}$ & $\xi_{H}^{d}$ & $\xi_{H}^{\ell}$
\\[-3mm]
&&&
\\
\hline
&&& \\[-3.5mm]
 Type-I~\, & $\sin\!\alpha/\!\sin\!\beta$
 & $\sin\!{\alpha}/\!\sin\!{\beta}$
 & $\sin\!{\alpha}/\!\sin\!{\beta}$
\\[1mm]
 Type-II & $\sin\!\alpha/\!\sin\!\beta$
 & $\cos\!{\alpha}/\!\cos\!{\beta}$
 & $\cos\!{\alpha}/\!\cos\!{\beta}$
\\[1mm]
\hline\hline
 \end{tabular}
\label{tab:1}
\vspace*{3mm}
\end{table}
\begin{figure}[t]
\vspace*{-4mm}
\begin{centering}
\includegraphics[width=8.0cm,height=6.5cm]{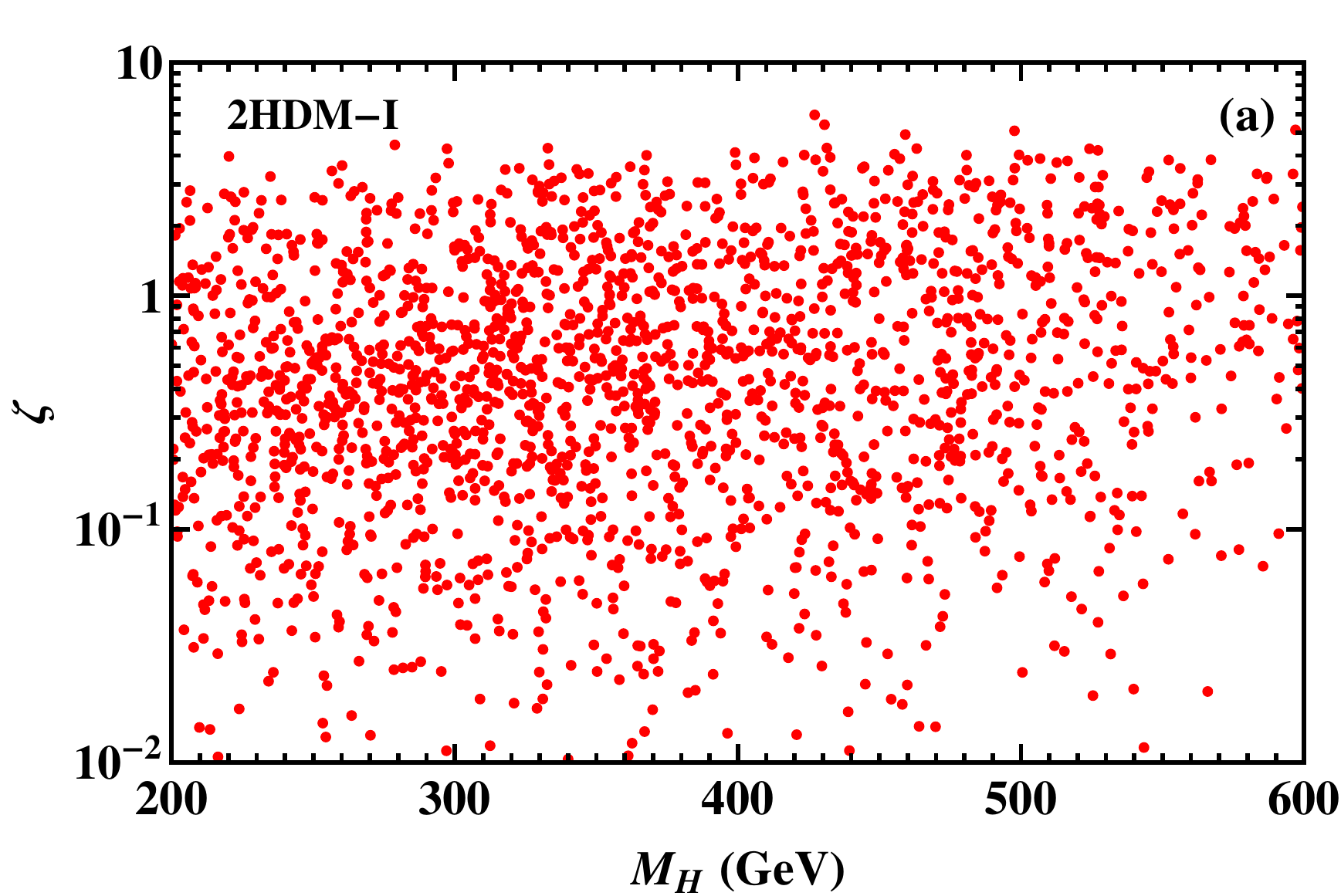}
\includegraphics[width=8.0cm,height=6.5cm]{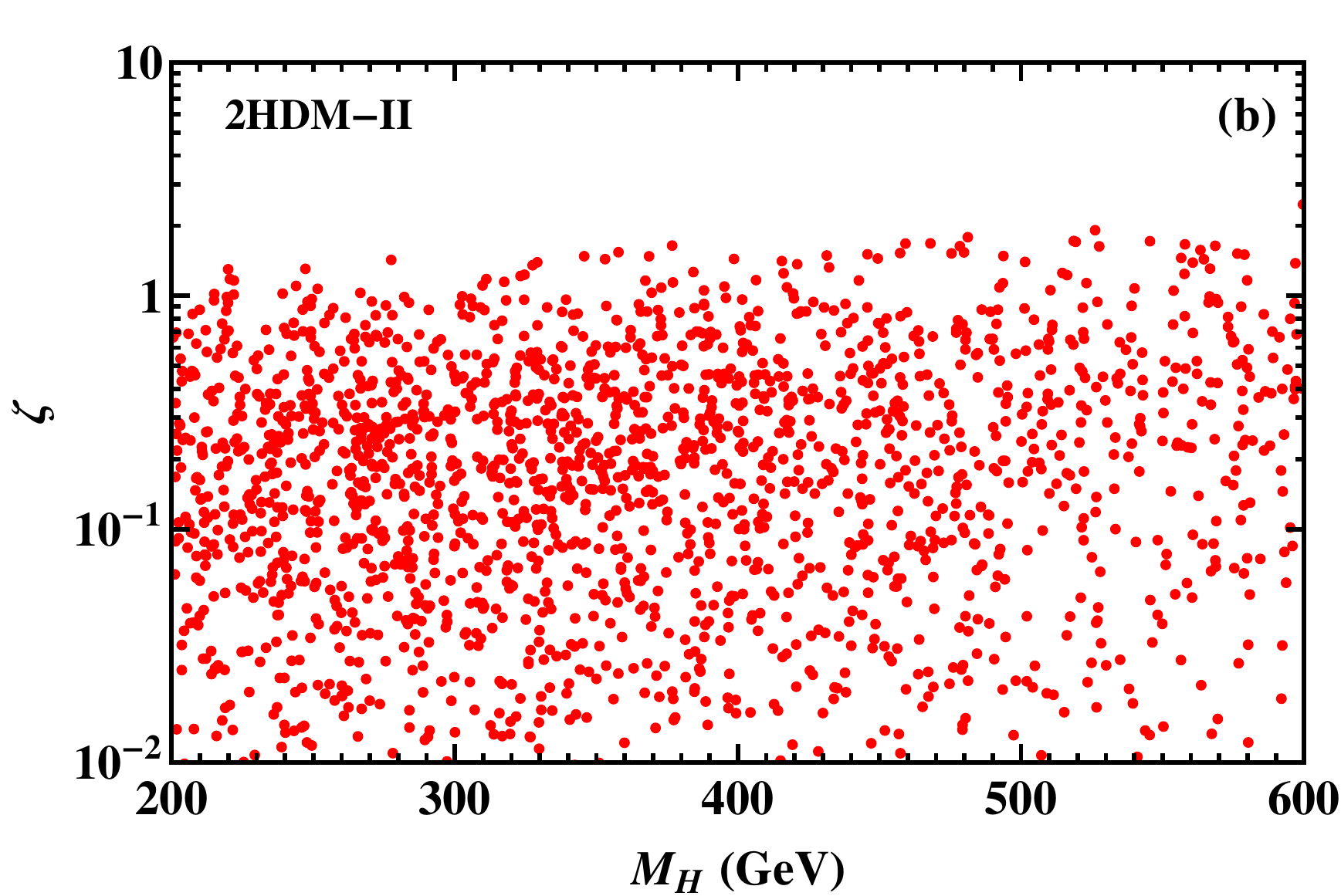}
\caption{Parameter space in $\,M_H^{}-\zeta\,$ plane for
2HDM-I [plot-(a)] and 2HDM-II [plot-(b)],
where the red dots present the viable points obeying the consistency requirement of
the Higgs potential as explained in the text.
}
\label{fig:1}
\end{centering}
\end{figure}

Inspecting the Higgs potential (\ref{eq:2HDM_potential}), we derive the scalar coupling
of trilinear vertex $\,Hhh$\,,
\begin{eqnarray}
\label{eq:Hhh_coup1}
G_{Hhh}^{}\,&=&\, \frac{\,\cos(\beta\!-\!\alpha)\,}{v}
\left[\(3M_{A}^{2} - M_{H}^{2} - 2M_{h}^{2} + 3\lambda_{5}^{} v^{2}\)
 \( \cos\!2(\beta\!-\!\alpha)-\frac{\,\sin\!2(\beta\!-\!\alpha)\,}{\tan\!{2\beta}}
 \) -M_{A}^{2}-\lambda_5^{}v^{2}
\right]
\nn\\
 &=&  \, \frac{\,\cos(\beta\!-\!\alpha)\,}{v}
\left[\(\frac{6 M_{\!12}^{2}}{\sin\! 2\beta} - M_{H}^{2} - 2M_{h}^{2} \)
 \( \cos\!2(\beta\!-\!\alpha)-\frac{\,\sin\!2(\beta\!-\!\alpha)\,}{\tan\!{2\beta}}
 \) -\frac{2M_{12}^{2}}{\,\sin\!2\beta\,}
\right] ,
\end{eqnarray}
where in the second step we have used the relation
$\,M_A^2 +\lambda_5^{}v^{2} =2M_{\!12}^{2} /\!\sin\!2\beta$\,.\,
In the SM, the cubic Higgs coupling $\,G_{hhh}^{\text{sm}}=-3M_h^2/v\,$.\,
We define a coupling ratio,
$\,\zeta = G_{Hhh}^{}/G_{hhh}^{\text{sm}}\,$,\,
which characterizes the relative strength of the $Hhh$ coupling as compared to the
$\,h^3\,$ Higgs coupling of the SM.  Under alignment limit $\,\cos(\beta-\alpha)\to 0\,$,\,
the trilinear scalar coupling (\ref{eq:Hhh_coup1})
takes the asymptotical form,
\beqa
\label{eq:eta}
\zeta ~=\, \frac{\,G_{Hhh}^{}\,}{G_{hhh}^{\text{sm}}} ~=~
\frac{\,({8M_{\!12}^{2} /\!\sin\! 2\beta}  -\! M_{H}^{2} \!-\!2M_{h}^{2})\,}{3M_h^2}
{\cos(\beta\!-\!\alpha)}
+\mO(\cos^2(\beta\!-\!\alpha))  \,.
\eeqa

In Fig.\,\ref{fig:1}, we explore the parameter space of the Higgs potential
(\ref{eq:2HDM_potential}) in the $\,M_H^{}-\zeta\,$ plane.
For $\,\zeta > 1\,$,\, we expect that the decay branching fraction $\,\text{Br}[H\to hh]\,$
and the production cross section $\,\sigma[gg\to H\to hh]\,$
will be enhanced by the factor $\,\zeta^2\,$.\,
In Fig.\,\ref{fig:1}, the red points present the viable parameter space
consistent with vacuum stability, unitarity and perturbativity bounds of the Higgs potential
\cite{Branco:2011iw}.
We also take into account the $3\sigma$ constraints from
the current Higgs global fit (cf.\ Sec.\,\ref{sec:4}).
The electroweak precision data also constrain the parameter space of the 2HDM.
It was found that in the 2HDM the charged Higgs mass satisfies,
$-600\,\GeV < M_{H\pm}^{} -M_{3}^{} <100\,\GeV$ and
$ M_{H\pm}^{} > 250\GeV$ \cite{Funk:2011ad},
where $\,M_{3}^{}\,$ is the mass of the heaviest neutral scalar.
In the case with exact $\,\ZZ_{2}^{}$ ($M_{12}^{}=0$),
the potential could be valid up to the scale $\sim\!10\TeV$ \cite{Chakrabarty:2014aya},
while for the present case of a softly broken $\,\ZZ_{2}^{}$,
the bound is much more relaxed, and the theory can be valid up to the Planck scale.
For the analysis of Fig.\,\ref{fig:1},
we have scanned the parameter space in the following range,
$\,\tanb \in [1,\,10]$,\, $\,\cosba\in [-0.6,\,0.6]$,\,
$\,M_{12}^{2} \in [-200^2,\,200^2]$\,GeV$^2$,\,
$\,M_H^{} \in [200,\, 600]$\,GeV, and
$\,M_A^{},M_{H\pm}^{}\in [300,\,2000]$\,GeV.\,
In the following analysis, we will consider the same range of
the 2HDM parameter space unless specified otherwise.

\vspace*{2mm}
\subsection{{\bf Heavier Higgs Boson $\,\mathbf{H}^{\bf 0}$: Decays and Production}}
\vspace*{2mm}

\begin{figure}
\vspace*{-3mm}
\begin{centering}
\includegraphics[width=8.0cm,height=6.5cm]{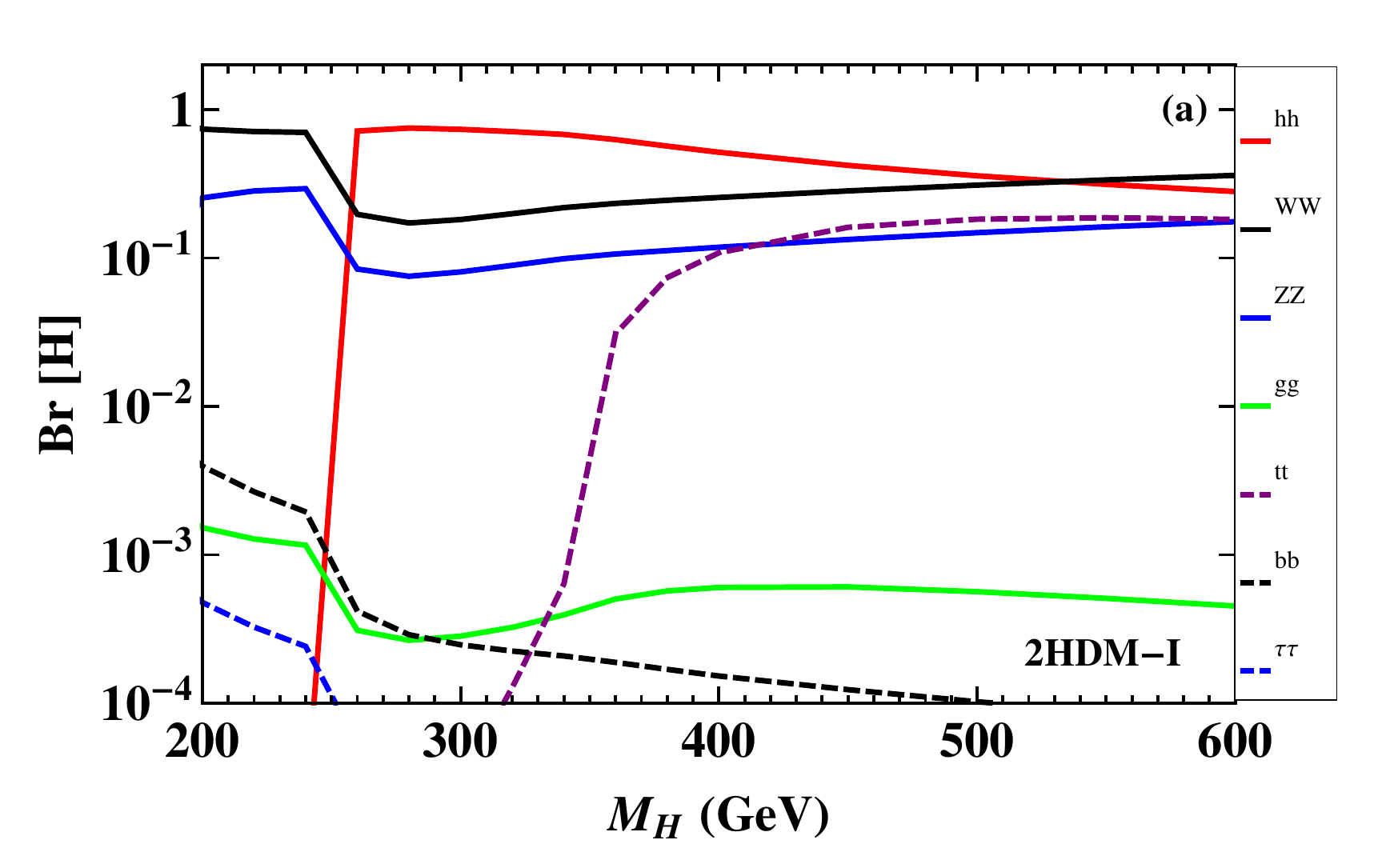}
\includegraphics[width=8.0cm,height=6.5cm]{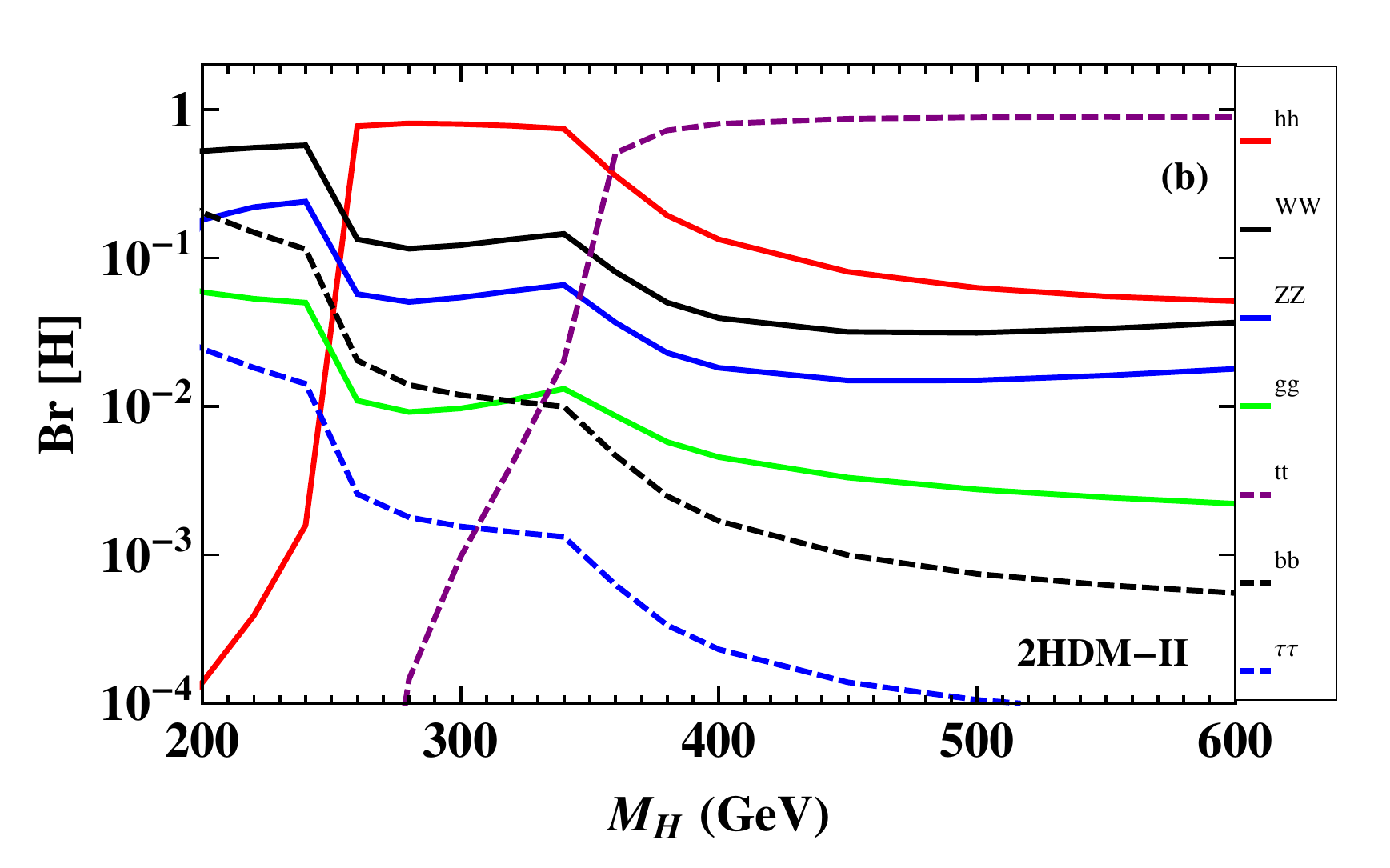}
\vspace*{-3mm}
\caption{Decay branching fractions of the heavier Higgs state \,$H^0$\,
for 2HDM-I [plot-(a)] and 2HDM-II [plot-(b)].\,
For illustration, we set $\,\tanb =1\,$ and
$\,(M_A^{},\,M_{\!12}^2)=(500\GeV,\,-(180\GeV)^2)$\,
for both plots. We also input $\,\cosba =0.4$\, for plot-(a) and
$\,\cosba =0.1$\, for plot-(b).
}
\label{fig:2}
\label{fig:2-HBr}
\end{centering}
\end{figure}

Let us consider the decay modes of the heavier neutral Higgs boson $H^0$.\,
It is straightforward to infer the tree-level decay width for $\,M_H^{}> 2M_h^{}\,$,
\beqa
\Gamma[H\to hh] ~=~
\frac{\,9\zeta^2M_h^4\,}{\,32\pi v^2M_H^{}\,}\sqrt{1-\frac{4M_h^2}{M_H^2}\,}\,.
\eeqa
For $\,M_H^{}\leqq 2M_h^{}\,$, we will include the off-shell decay
$\,H\to hh^*\,$ with $\,h^*\to f\bar{f},\,gg,\,\gamma\gamma\,$,\, etc, where $\,f\,$
denotes the light fermions except top quark.
For the decay modes $\,H\to VV,\,f\bar{f}\,$,\, we have
$\,\Gamma[H\to VV]/\Gamma[H\to VV]_{\text{sm}}=\cos^2(\beta-\alpha)\,$
and
$\,\Gamma[H\to f\bar{f}]/\Gamma[H\to f\bar{f}]_{\text{sm}}=(\xi_H^f)^2\,$.\,
(Here, the subscript ``sm" denotes the ``standard model" with a reference
Higgs boson $\,H\,$ which has the same mass as $\,H\,$ in the 2HDM.)\,
For the decay channel $\,H\to gg$\,,\, we can express the partial width relative to the SM value,
$\,\Gamma[H\to gg]/\Gamma[H\to gg]_{\text{sm}}
 =|\sum_{f=t,b}\xi_H^fA_{1/2}^H(\tau_f^{})/A_{1/2}^H(\tau_t^{})|^2\,$,\,
where $\,\tau_f^{}=M_H^2/(4m_f^2)\,$
and the function $\,A_{1/2}^H(\tau_f^{})\,$ is the standard formula \cite{Branco:2011iw}\cite{Djouadi:2005gi}.
The decay branching ratio of $\,H\to \gamma\gamma\,$ is practically negligible
for $\,M_H\gtrsim 200\,$GeV.\,
In Fig.\,\ref{fig:2}, we present the decay branching fractions
of the heavier Higgs boson $\,H\,$ for both 2HDM-I [plot-(a)] and 2HDM-II [plot-(b)].
For illustration, we input $\,\tanb =1\,$ and
$\,(M_A^{},\,M_{\!12}^2)=(500\GeV,\,-(180\GeV)^2)$\,
for both plots. We also set $\,\cosba =0.4$\, for plot-(a) and
$\,\cosba =0.1$\, for plot-(b).
We see that for $\,M_H^{}<250$\,GeV,\, the dominant decay channels are
$\,H\to ZZ, WW$,\, and for $\,250\,\GeV < M_{H}^{} <350$\,GeV,\,
the major decay channels include $\,H\to ZZ, WW, hh$\, since the $\,H\to hh\,$ channel opens up.
For $\,M_H^{}>350\GeV$, the $\,H\to t\bar{t}$\, channel is further opened, and
will become dominant in 2HDM-II when $\,\cosba\,$ takes values around
the alignment limit as shown in Fig.\,\ref{fig:2}(b).
But this situation can change when $\,\cosba\,$ becomes larger and falls into the allowed region
which separates from the alignment region
(cf.\ Fig.\,\ref{fig:9} in Sec.\,\ref{sec:4}).

From Eq.\,(\ref{eq:H_Yukawa}) and Tabel\,\ref{tab:1}, we see that the Yukawa coupling of the
heavier Higgs boson $H$ with $t\bar{t}$\, has a scale factor
$\,\xi_{H}^{t}=\sin\!\alpha^{}/\!\sin\!\beta^{}$\, relative to the SM Higgs Yukawa coupling.
The major LHC production channel is the gluon fusion process $\,gg\to H\,$.\,
Other production processes include the vector boson fusion $\,pp\to H qq'\,$,\,
the vector boson associated production $\,pp\to HV$,\, and the top associated production
$\,gg\to Ht\bar{t}\,$.\,
The gluon fusion production cross section of $\,H\,$ can be obtained from the corresponding
SM cross section with a rescaling by $\,H\to gg\,$ partial width,
\beqa
\label{eq:ggH_xsec}
\sigma[gg \!\to\! H] ~=\,
\({\Gamma[H\!\to\! gg]}/{\Gamma[H\!\to\! gg]_{\text{sm}}^{}}\)
\sigma[gg\!\to\! H]_{\text{sm}}^{} \,,
\eeqa
where we will include all NLO QCD corrections to the gluon fusion cross section
as done in the SM case\,\cite{NLO}.
We note that for 2HDM-I, Table\,\ref{tab:1} shows that the $H$ Yukawa couplings
with top and bottom quarks have the same structure as in the SM, so the $H$ production
cross section $\,\sigma[gg\!\to\! H]\,$ differs from the SM by a simple rescaling factor
$\,(\sin\!\al /\!\sin\!\beta)^2$.\,
For the 2HDM-II, we see that the $H$ coupling to $b$ quarks differs from that
of $t$ quarks by a factor of $\,\tanb /\!\tan\!\al\,$,\, which can enhance the $b$-loop
contribution to $\,gg\!\to\! H\,$ production for large $\,\tanb\,$ region.
Hence, the general relation \eqref{eq:ggH_xsec} should be used.
The uncertainty of the gluon fusion cross section is about $10\%$ over the mass-range
$\,M_H^{}= 250-600\,$GeV \cite{NLO}, which is roughly the total uncertainty
of signal and background calculations.

For the inclusive $\,H\,$ production, we include the gluon fusion $\,gg\to H,\,$ and
$b$-related processes $b\bar{b}\to H$,\, $gb\,(g\bar{b})\to H b\,(H\bar{b})$,\, and
$gg\,(q\bar{q})\to Hb\bar{b}\,$.\,
The production cross sections for these $b$-related processes are derived by rescaling
a factor of $\,(\xi_{H}^{d})^{2}$\, from their corresponding SM productions
with the same Higgs mass. So we have the total inclusive cross section of
$\,pp \to HX\,$ for the 2HDM,
\begin{eqnarray}
\label{eq:CX-HX}
\sigma[pp\to HX] &\!\!=\!\!&
\({\Gamma[H\!\to\! gg]}/{\Gamma[H\!\to\! gg]_{\text{sm}}^{}}\)
\sigma[pp(gg)\to H]_{\text{sm}}^{}
\nn\\
&&
+\,(\xi_{H}^{d})^{2}\left\{\sigma[pp(b\bar{b})\to H]_{\text{sm}}^{}
    \!+ \sigma[pp(gb,g\bar{b})\to Hb,H\bar{b}]_{\text{sm}}^{}
    \!+ \sigma[pp(gg,q\bar{q})\to Hb\bar{b}]_{\text{sm}}^{} \right\} .
~~~~
\end{eqnarray}
%


%
\begin{figure}[t]
\vspace*{-7mm}
\begin{centering}
\hspace*{-5mm}
\includegraphics[width=7.5cm,height=6.3cm]{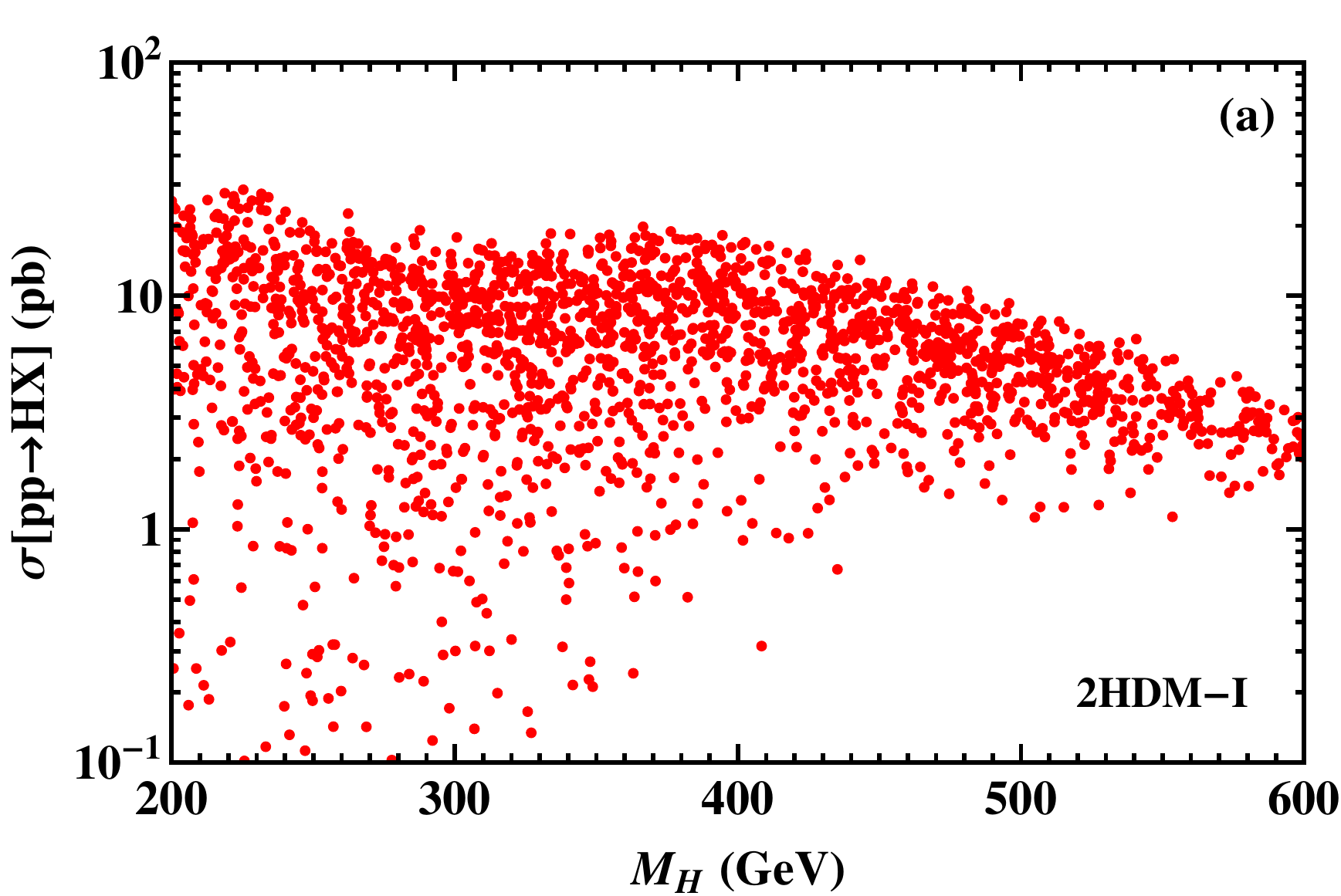}
\hspace*{4mm}
\includegraphics[width=7.5cm,height=6.3cm]{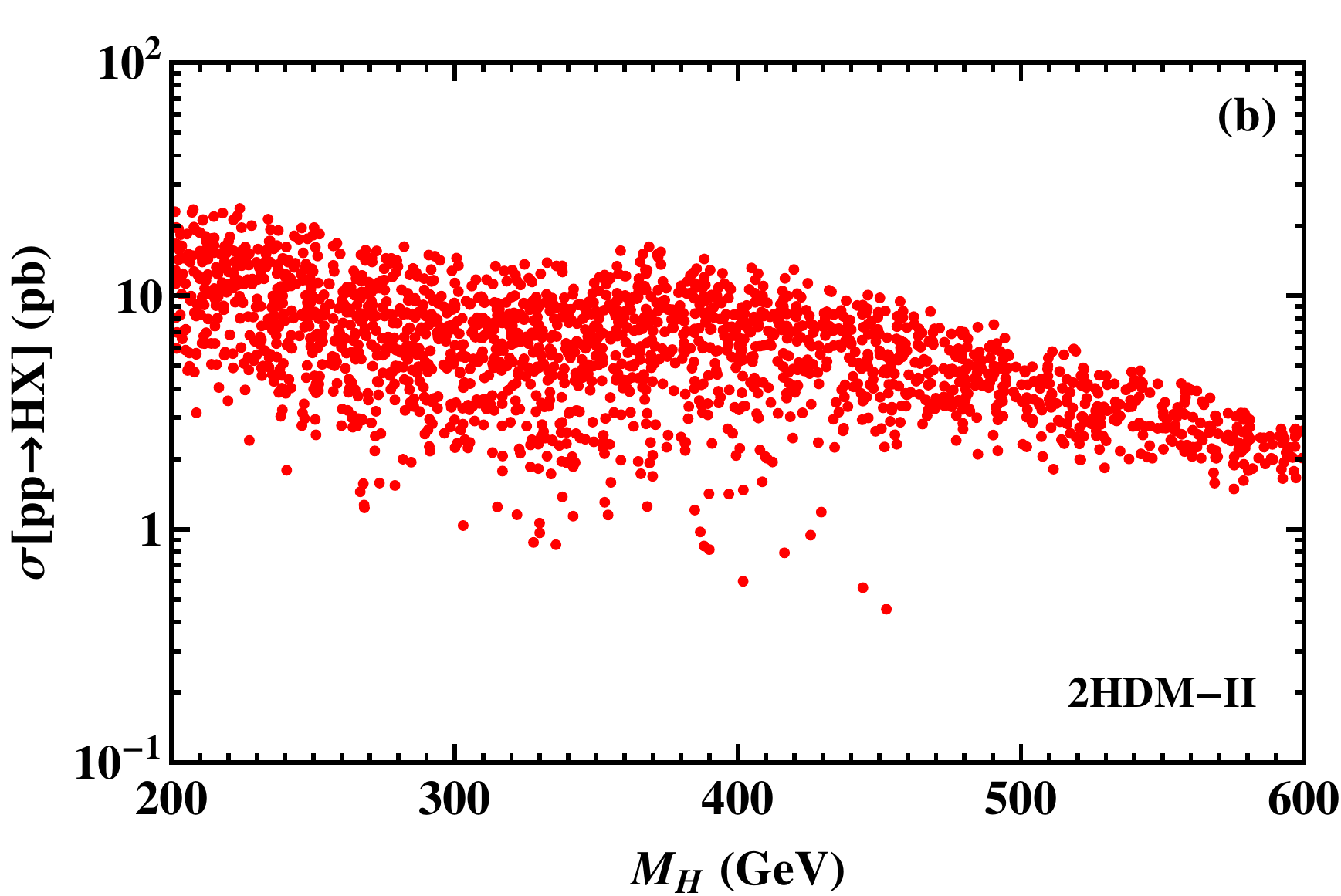}
\\[-0.2mm]
\includegraphics[width=8.2cm,height=6.1cm]{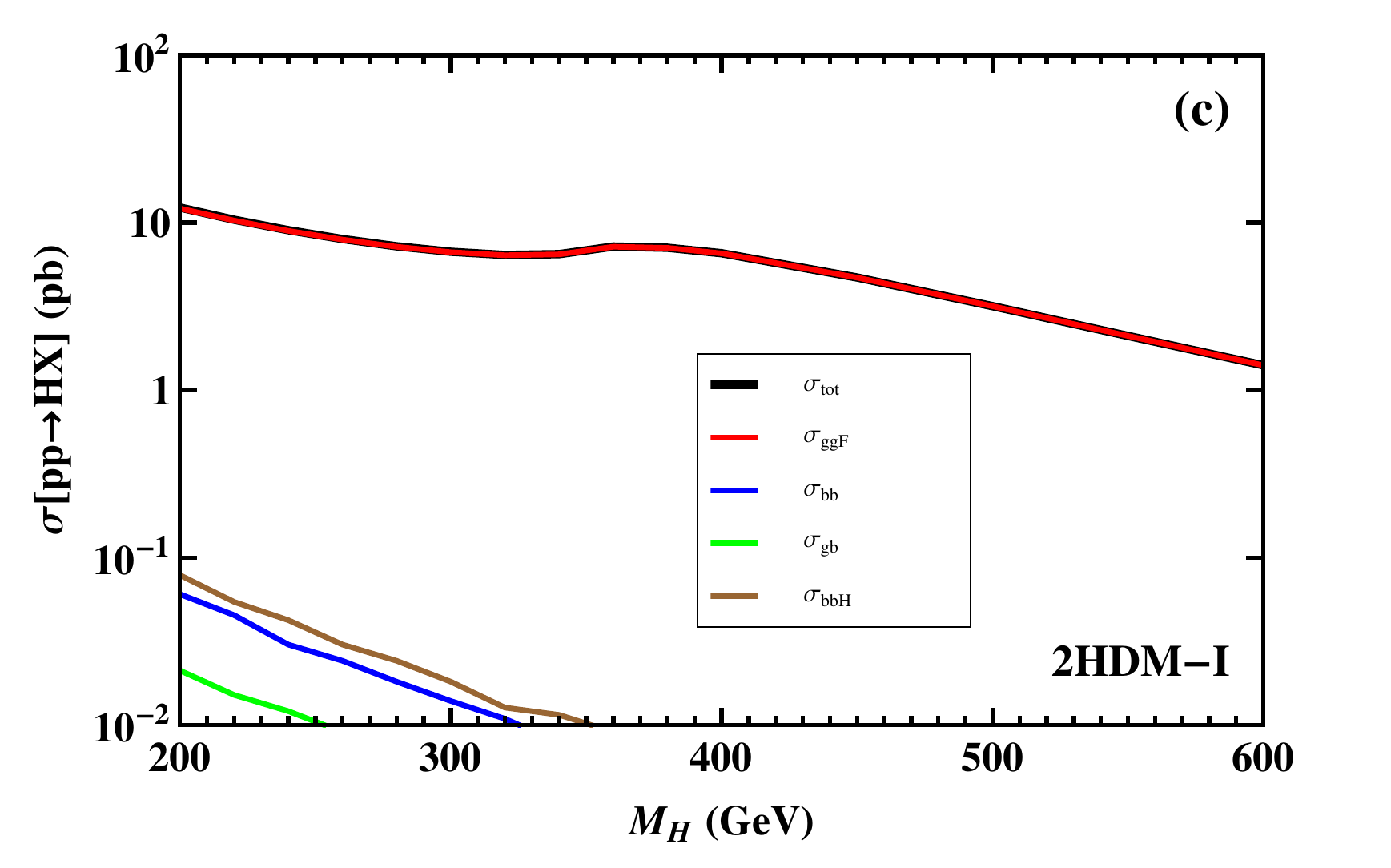}
\hspace*{-2mm}
\includegraphics[width=8.2cm,height=6.1cm]{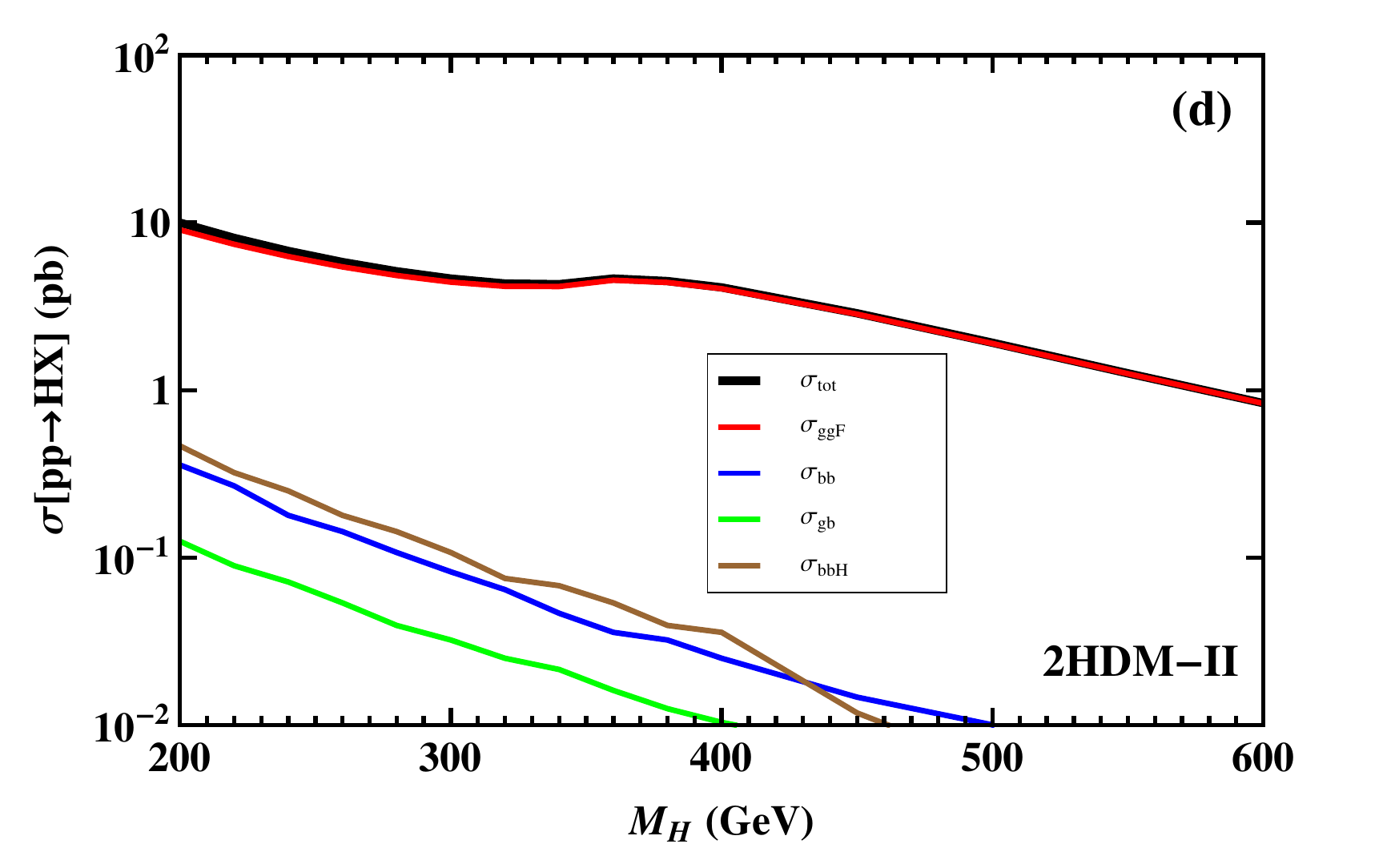}
\vspace*{-3mm}
\caption{Inclusive $H$ production cross section via $\,pp\to HX\,$ process at the LHC\,(14\,TeV),
for 2HDM-I [plot-(a)] and 2HDM-II [plot-(b)] with $\,\tanb \in [1,\,10]$.\,
All the red points satisfy the requirements of stability, perturbativity and unitarity,
as well as the $3\sigma$ constraint by the current Higgs global fit.
The cross section of inclusive $H$ production $\,pp\to HX\,$ contains four sub-channels
from $\,gg\to H$,\, $b\bar{b}\to H$,\, $gb\,(g\bar{b})\to H b\,(H\bar{b})$,\, and
$\,gg\,(q\bar{q})\to Hb\bar{b}\,$.\,
In plot-(c) and plot-(d), we present the sub-channel contributions to the inclusive
cross section $\,\sigma [pp\to HX]\,$ for 2HDM-I and 2HDM-II, respectively,
where we set sample inputs,
$\,\tan\beta=2\,$ and $\,\cos(\beta-\alpha)=-0.3\,(-0.1)$\, for 2HDM-I (2HDM-II).
In plots (c)-(d), the red curve ($gg\to HX$ contribution) and the black curve
(summed total contribution) fully overlap because the $\,gg\to HX\,$ channel
dominates the inclusive cross section in the low $\tan\beta$ region.
}
\label{fig:2-pptoHX}
\label{fig:3}
\end{centering}
\vspace*{-3mm}
\end{figure}

\begin{figure}[t]
\begin{centering}
\includegraphics[width=8.0cm,height=6.5cm]{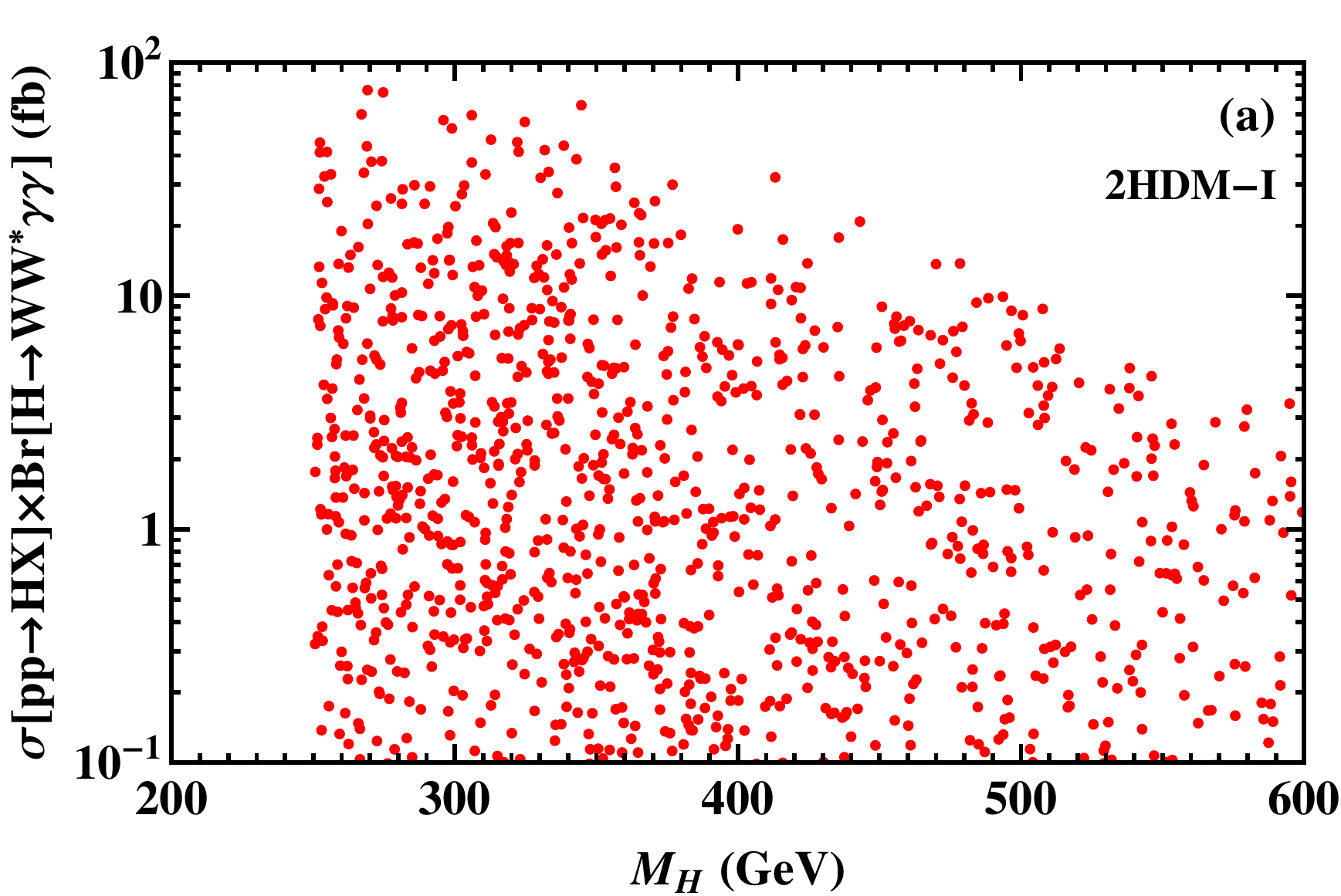}
\includegraphics[width=8.0cm,height=6.5cm]{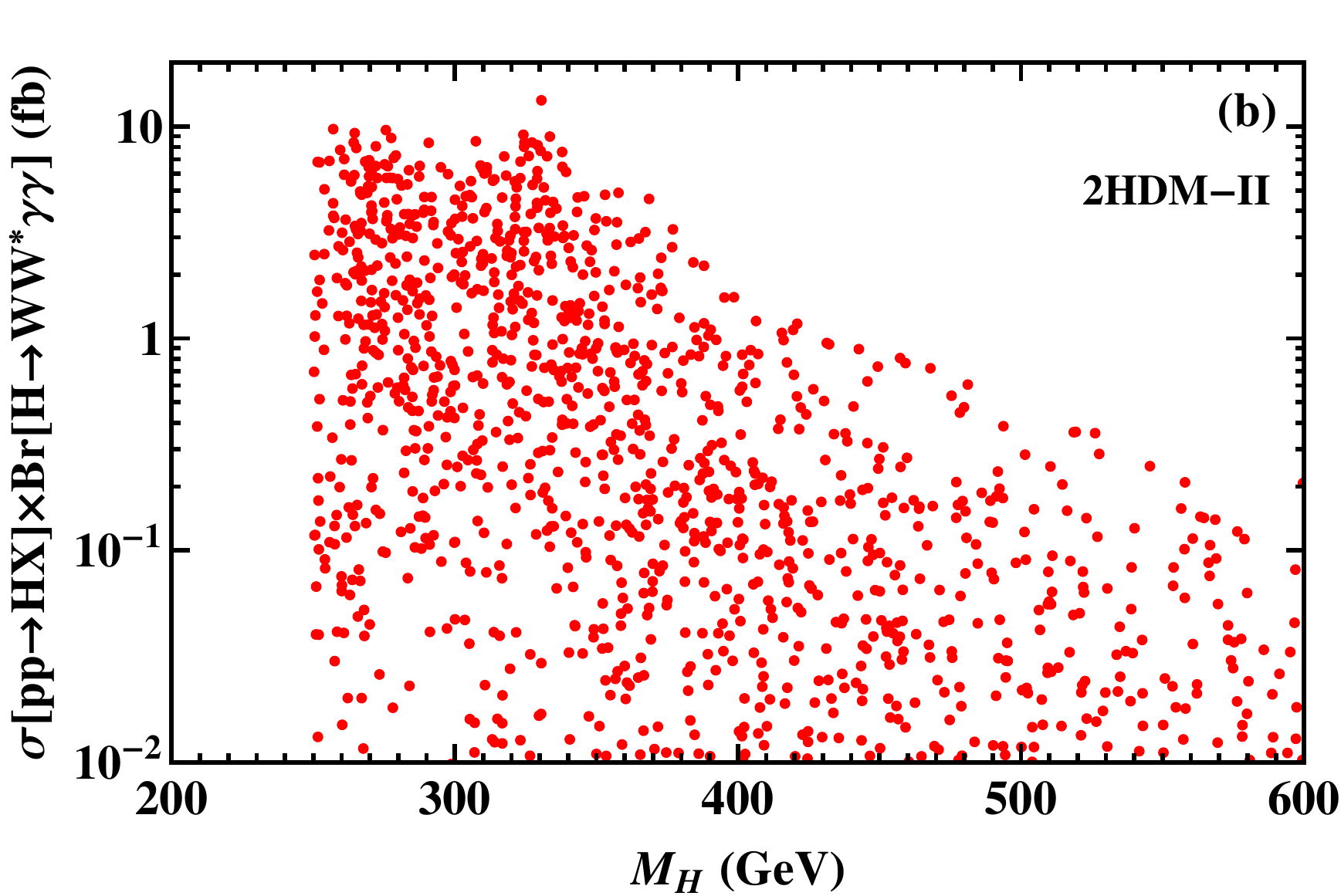}
\vspace*{-1.5mm}
\caption{LHC signal cross section
$\,\sigma(pp\!\to\! HX)\!\times\!\text{Br}(H\!\to\! hh\!\to\! WW^*\gamma\gamma)$\,
in the 2HDM with $\,\tanb \in [1,\,10]$.\,
Plots (a) and (b) present the results for 2HDM-I and 2HDM-II, respectively.}
\label{fig:3-Sigma-pptoHtowwaa}
\label{fig:4}
\end{centering}
\end{figure}

We present the inclusive $\,H\,$ production rate for 2HDM Type-I and Type-II
in Fig.\,\ref{fig:3}(a)-(b).
Multiplying the production cross section with decay branching fraction
$\text{Br}(H\!\to\! hh\!\to\! WW^*\gamma\gamma)$,\,
we compute the signal rate in the channel\footnote{Our analysis of the
production rate of $\,gg\to H\to hh\,$ in the 2HDM is consistent with
the recent study\,\cite{2HDM-new}. We thank Yun Jiang and J\'{e}r\'{e}my Bernon
for providing data points of their calculation for numerical comparison.}
$\,pp\!\to\! HX\,$ with $\,H\!\to\! hh \!\to\! WW^*\gamma\gamma$\,.\,
We summarize our results in Fig.\,\ref{fig:3-Sigma-pptoHtowwaa}
for 2HDM-I and 2HDM-II, respectively.
In Fig.\,\ref{fig:3}(a)-(b) and Fig.\,\ref{fig:4}(a)-(b),
we have scanned the same 2HDM parameter space as in Fig.\,\ref{fig:1}.
The signal process is depicted by the left diagram of Fig.\,\ref{fig:5}.
From Fig.\,\ref{fig:3-Sigma-pptoHtowwaa}, we see that the cross section
$\,\sigma(pp\!\to\! HX)\times\text{Br}(H\!\to\! hh\!\to\! WW^*\gamma\gamma)$\,
can be as large as about $70$\,fb for 2HDM-I; while for 2HDM-II,
this cross section can reach about $10$\,fb for $\,M_H^{}\lesssim 340$\,GeV\,.

For comparison, we show the individual contributions of each
sub-channel to the total inclusive cross section $\,\sigma[pp\to HX]\,$
in Fig.\,\ref{fig:2-pptoHX}(c)-(d).
For illustrations,  we set sample parameter inputs,
$\,\tan\beta=2\,$ and $\,\cos(\beta-\alpha)=-0.3$\, for 2HDM-I, and
$\,\tan\beta=2\,$ and $\,\cos(\beta-\alpha)=-0.1\,$ for 2HDM-II.
In plots (c)-(d), the red curve ($gg\to H$ contribution) fully overlaps
the black curve (summed total contribution).
This is because the gluon fusion channel $\,gg\to H\,$
dominates the inclusive production cross section for low $\,\tanb\,$ region of the 2HDM.
In general, Table\,\ref{tab:1} shows that for 2HDM-I, the $H$ Yukawa couplings
($\,\xi^u_H =\xi^d_H = \sin\alpha/\!\sin\beta\,$)  are rather insensitive to $\tanb$\,.\,
Hence, in 2HDM-I the gluon fusion actually dominates the $\,H$\, production
over full range of $\,\tanb\geqq 1$,\, and the contributions of $b$-related sub-channels
are always negligible.  For 2HDM-II, the (up-type) $H$ Yukawa coupling
$\,\xi^u_H= \sin\alpha/\!\sin\beta\,$ is the same as 2HDM-I, and the down-type Yukawa coupling
$\,\xi^d_H\propto 1/\!\cos\beta = \tanb/\!\sin\beta\,$ is enhanced
by a $\,\tanb\,$ factor relative to $\,\xi^u_H\,$.\, We find that
for small $\,\tanb\lesssim 3\,$,\, the gluon fusion channel still dominates
the inclusive $\,H$\, production in 2HDM-II, and its cross section is larger than
other $b$-related channels by a factor of $\,O(10\!-\!100)\,$ for $M_H^{}\geqq 300$\,GeV.\,
The analysis of 2HDM-II in Sec.\,\ref{sec:4} also concerns the small $\,\tanb\,$ region
[cf.\ Fig.\,\ref{fig:9}(b)(d)]. Hence, in the following Sec.\,\ref{sec:3}--\ref{sec:4},
we will focus our analysis on the Higgs production from gluon fusion channel,
$pp(gg)\to H\to hh\to WW^*\gaga$\,.

\vspace*{2mm}
\section{Higgs Signal and Background Simulations}
\label{sec:3}
\vspace*{1.5mm}

\begin{figure}[b]
\vspace*{-3mm}
\begin{centering}
\includegraphics[width=7cm,height=3.8cm]{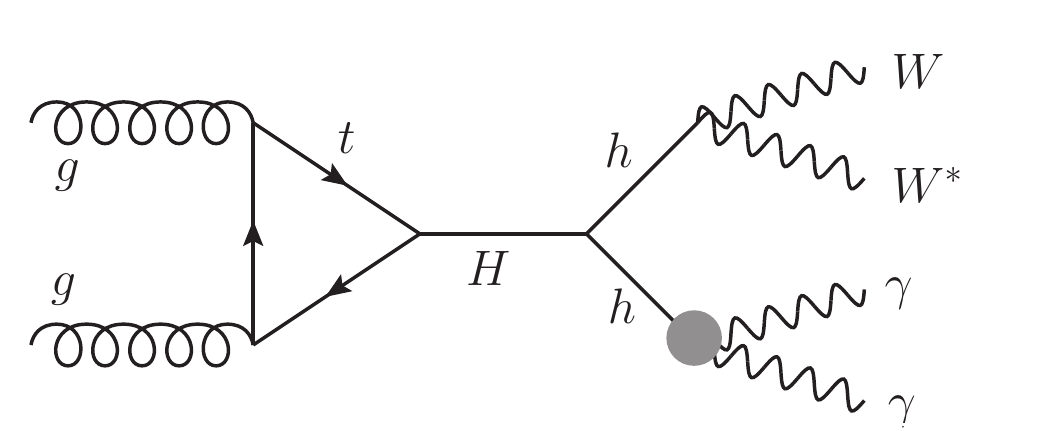}
\includegraphics[width=7cm,height=4.0cm]{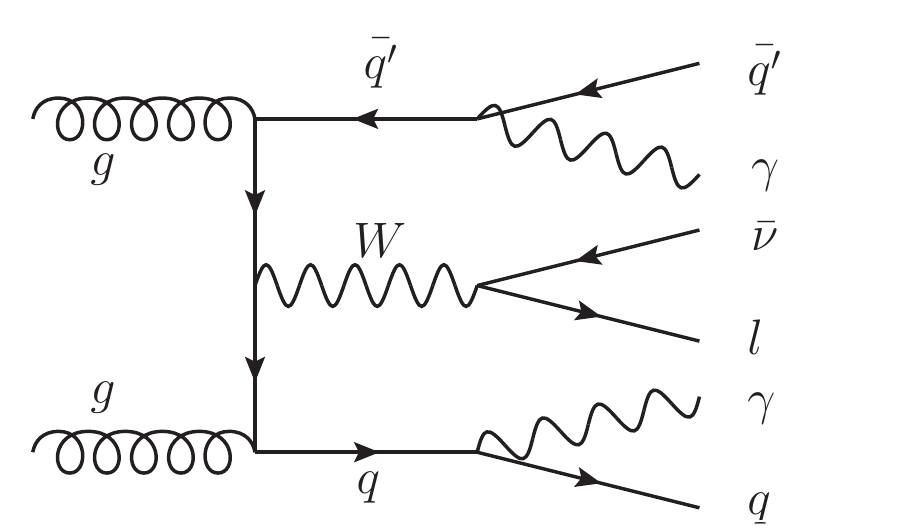}
\caption{LHC production process $\,gg\to WW\gamma\gamma$\,.\,
The left diagram shows the signal production via $\,gg\to H\to hh\to WW\gamma\gamma\,$,\,
and the right diagram illustrates an irreducible background process
$\,gg \to q\bar{q'}\ell\bar{\nu}\gaga\,$.
}
\label{fig:4-pptoHtowwaa}
\label{fig:5}
\end{centering}
\end{figure}

In this section, we compute the Higgs signals and backgrounds at the LHC\,(14\,TeV).
We perform systematical simulations by using {MadGraph5} package\,\cite{Alwall:2014hca}
for the process, $\,pp(gg) \!\to\! H \!\to\! hh \!\to\! WW^*\gamma\gamma$,\,
via gluon fusion channel. The parton-level Higgs production cross section
$\sigma(gg\to H)$ is derived from Eq.\,\eqref{eq:ggH_xsec}, including NLO QCD corrections.
We illustrate the signal Feynman diagram by the left plot of Fig.\,\ref{fig:4-pptoHtowwaa}.
For signal process, we generate the model file using {FeynRules}\,\cite{Alloul:2013bka},
containing $\,Hhh\,$ vertex and the effective $\,ggH\,$ vertex.
We compute signal and background events using {MadGraph5/MadEvent} \cite{Alwall:2014hca}.
Then, we apply {Pythia}\,\cite{Sjostrand:2006za} to simulate hadronization of partons and
adopt {Delphes}\,\cite{deFavereau:2013fsa} to perform detector simulations.

For the final state $WW$ decays, we will study both the pure leptonic mode
$\,WW\to \ell\nu\ell\nu\,$ and the semi-leptonic mode
$\,WW\to q\bar{q}'\ell\nu$.\,
The $W$ decay branching fractions to $\,e\nu\,$ and $\,\mu\nu\,$ equal
10.8\% and 10.6\%, respectively,
while that of $\,W\to\tau\nu\,$ is about 11.3\% \cite{Beringer:1900zz}.
The dijet branching ratio of $\,W\to q\bar{q}'\,$ equals 67.6\% \cite{Beringer:1900zz}.
Thus, the inclusion of semi-leptonic mode will be beneficial.
Since $\,\tau\,$ leptons can decay into $\,e,\mu$\,,\,
the detected final state $\,e,\mu$\,  will include those from the $\tau$ decays.
For $\,M_h=125\,$GeV,\, the branching fraction of $\,h\to\ga\ga\,$ in the SM equals
$\,2.3 \times 10^{-3}\,$ \cite{Djouadi:2005gi}.
In the following, we will first present the analyses for $\,M_H^{}=300\,$GeV
in Sec.\,\ref{sec:3.1}--\ref{sec:3.2},
and then for heavier masses $\,M_H^{}=400,600\,$GeV in Sec.\,\ref{sec:3.3}.

\vspace*{2mm}
\subsection{{\bf Pure Leptonic Channel:}
${hh\to\! WW^*\gaga\! \to \ell\nu \ell\nu\ga\ga}$}
\label{sec:lvlvaa}
\label{sec:3.1}
\vspace*{3mm}

For pure leptonic channel, we have $\,hh\to WW^*\gaga\to \ell\nu\ell\nu\gaga\,$.\,
Although this channel has an event rate about two orders of magnitude lower than that of
$\,hh\to b\bar{b}\gamma\gamma$\, mode,  it has much cleaner background as compared to
$\,b\bar{b}\gamma\gamma$\, final state.
After imposing simple cuts, we find that the backgrounds can be substantially reduced.
We follow the ATLAS procedure for event selections.
To discriminate the Higgs signal from backgrounds, we set up preliminary event selection
by requiring two leptons (electron or muon) and at least two photons in the final state,
\beqa
\label{eq:selection}
n_{\ell}^{} = 2 \,,
\qquad
n_{\gamma}^{} \geqq 2 \,.
\eeqa

In the first step of event analysis, we need to prevent the potential double-counting, i.e.,
the reconstructed objects are required to have a minimal spatial separation \cite{ATLAS2013}.
The two leading photons are always kept, but we impose the following criteria \cite{ATLAS2013}:
(i) electrons overlapping with one of those photons within a cone
$\,\Delta R (e,\gamma) <0.4$\, are rejected;
(ii) jets within $\,\Delta R (\text{jet},e) <0.2$\, or
     $\,\Delta R (\textrm{jet},\gamma) <0.4$\, are rejected;
(iii) muons within a cone of $\,\Delta R (\mu, \textrm{jet}) <0.4$\, or
     $\Delta R (\mu,\gamma) <0.4$\, are rejected.
After this, we apply the basic cuts to take into account the detector conditions,
which are imposed as follows,
\begin{eqnarray}
\label{eq:basic-cut}
P_{T}^{}(\gamma), P_{T}^{}(q) > 25\,\mathrm{GeV}, \quad
P_T^{}(\ell) > 15\,\mathrm{GeV}, \quad
|\eta(\gamma)|, |\eta(q)|, |\eta (\ell)| < 2.5 \,.
\end{eqnarray}
\begin{table}[t]
\vspace*{-5mm}
\centering
\caption{Signal and background cross sections of
$\,pp\to WW^*\gaga\to \ell\nu\ell\nu\gamma\gamma$\, and
$\,pp\to WW^*\gaga\to q\bar{q}'\ell\nu\gamma\gamma$\,
processes at the LHC\,(14\,TeV) after each set of cuts.
The signal significance($Z_0$) is computed for the LHC\,(14\,TeV) runs with
300\,fb$^{-1}$ integrated luminosity. We input the heavier Higgs mass $\,M_H^{}=300\,$GeV,\,
and set the sample signal cross section as
$\,\sigma(pp\!\to\! H\!\to\! hh\!\to\! WW^*\gamma\gamma) = 5\,\mathrm{fb}$\,.
From the 3rd to 5th columns, we show the signals and backgrounds after imposing each set
of cuts. The ``Selection + Basic Cuts" are choosing according to
Eqs.\,\eqref{eq:selection}--\eqref{eq:basic-cut}.
In the pure leptonic mode, we impose the Final Cuts
$M_T^{}(\ell\ell\nu\nu), M(\ell\ell)$,\,
$M_T^{}(\ell\ell\nu\nu\gamma\gamma)$,\,
$\Delta\phi (\ell\ell)$,\, $\Delta R(\ell\ell)$,\, and $\Delta R(\gaga)$.\,
In the semi-leptonic mode, we add the Final Cuts
$P_T^{}(\gamma)$\, $\,M_T^{}(q\bar{q}'\!\ell\nu)$,\, and $\Delta R(\gaga)$.\,
}
\vspace*{-3mm}
\vspace{0.5cm}
\begin{tabular}{c||c|c|c|c}
\hline\hline
$\,pp\to \ell\nu\ell\nu\gamma\gamma$ & Sum
& Selection+Basic\,Cuts & $M_{\gamma\gamma}^{}, E\sla_{T}$
& Final Cuts
\tabularnewline
\hline
Signal\,(fb) & 0.525 & 0.0251 & 0.0214 & 0.0161
\tabularnewline
BG$[\ell\nu\ell\nu\gamma\gamma\!+\!\ell\ell\gaga]$ (fb)
& 153.3 & 0.937 & 0.00225 & 0.000215
\tabularnewline
BG$[t\bar{t}h]$ (fb)
& 0.0071 & 0.000493 & 0.000419 & 0.000076
\tabularnewline
BG$[Zh]$ (fb)
& 0.175 & 0.0331 & 0.00210 & 0.000078
\tabularnewline
BG$[hh]$ (fb)
& 0.00222 & 0.000132 & 0.000102 & 0.000062
\tabularnewline
BG[Total] (fb)
& 153.48 & 0.971 & 0.00488 & 0.00043
\tabularnewline
Significance($Z_0$) & 0.734 & 0.439 & 3.70 & 5.15
\tabularnewline
\hline
\hline
$\,pp\to q\bar{q}'\ell\nu\gamma\gamma$ &
Sum & Selection+Basic\,Cuts & $M_{\gamma\gamma}^{}$,\,$M_{qq}^{}$,\,$E\sla_T^{}$
& Final Cuts
\\
\hline
Signal (fb) & 2.2 & 0.124 & 0.0937 & 0.0749
\\
BG$[q\bar{q}'\!\ell\nu\gamma\gamma]$ (fb) & 31.59 & 0.580 & 0.0192 & 0.00912
\\
BG$[\ell\nu\gamma\gamma]$ (fb) & 143.3 & 0.0642 & 0.00349 & 0.00182
\\
BG$[Wh]$ (fb) &0.42 & 0.00509 & 0.00234 & 0.00140
\\
BG$[WWh]$ (fb) & 0.0023 & 0.000210 & 0.000104 & 0.000050
\\
BG[$t\bar{t}h$] (fb) & 0.0148 & 0.00163 & 0.000802 & 0.000420
\\
BG[$hh$] (fb) & 0.00462 & 0.000291 & 0.000160 & 0.000106
\\
BG[$th$] (fb) &  0.0129 & 0.000479 & 0.000186 & 0.000099
\\
BG[Total] (fb) & 175.35 & 0.652 & 0.0264 & 0.0130
\\
Significance($Z_0)$ & 2.87 & 2.59 & 7.29 & 7.47
\\
\hline
\hline
\end{tabular}
%
\label{tab:2}
\end{table}

\vspace*{2mm}

Next, we turn to the background analysis for pure leptonic mode.
Besides the $\,\ell\nu\ell\nu \gamma\gamma\,$ and $\,\ell\ell\gaga\,$ backgrounds,
there are additional reducible backgrounds from Higgs bremstrahlung,
vector boson fusion, and $t\bar{t}h$ production.
The cross section of the former two processes are fairly small and thus negligible for
the present study. The $t\bar{t}h$ associate production, with $\,t\bar{t}\to WWb\bar{b}\,$,
can be important because the diphoton invariant-mass cut does not effectively discriminate
the signal process. But, this background can be suppressed by imposing $b$-veto \cite{b-veto}.
The production cross section for $\,t\bar{t}h\,$ in the SM is
$\,\sigma(pp\rightarrow t\bar{t}h) = 0.6113$\,pb \cite{tth}.
The latest $b$-veto efficiency of ATLAS is,
$\,\epsilon(b\mathrm{-veto})=22\%$\, \cite{b-veto-latest}.
Thus, we estimate the cross section for this background process,
\begin{equation}
\sigma(pp\!\rightarrow\! t\bar{t}h \!\rightarrow\! \ell\nu\ell\nu \gamma\gamma)
\,=\, \sigma(pp\!\rightarrow\! t\bar{t}h) \times
\mathrm{Br}[W\!\rightarrow\!\ell\nu]^2\,
\mathrm{Br}[h\!\rightarrow\!\gamma\gamma]\, \epsilon(b\mathrm{-veto})^2
\simeq 7.28 \!\times\! 10^{-3} \,\mathrm{fb}\,,~~~~~
\end{equation}
where $\,W\!\rightarrow\!\ell\nu\,$ includes $\,\ell =e,\mu,\tau$.\,
We see that imposing the $b$-veto has largely suppressed the $\,t\bar{t}h$\, background.
We note that the $\,t\bar{t}h\,$ background is much smaller than the
$\,\ell\nu\ell\nu\gamma\gamma\,$ background before kinematic cuts,
while after all the kinematic cuts it could be non-negligible.
So we will include both for the present background analysis.

%
\begin{figure}[H]
\vspace*{-2mm}
\begin{centering}
\includegraphics[width=8.0cm,height=5.8cm]{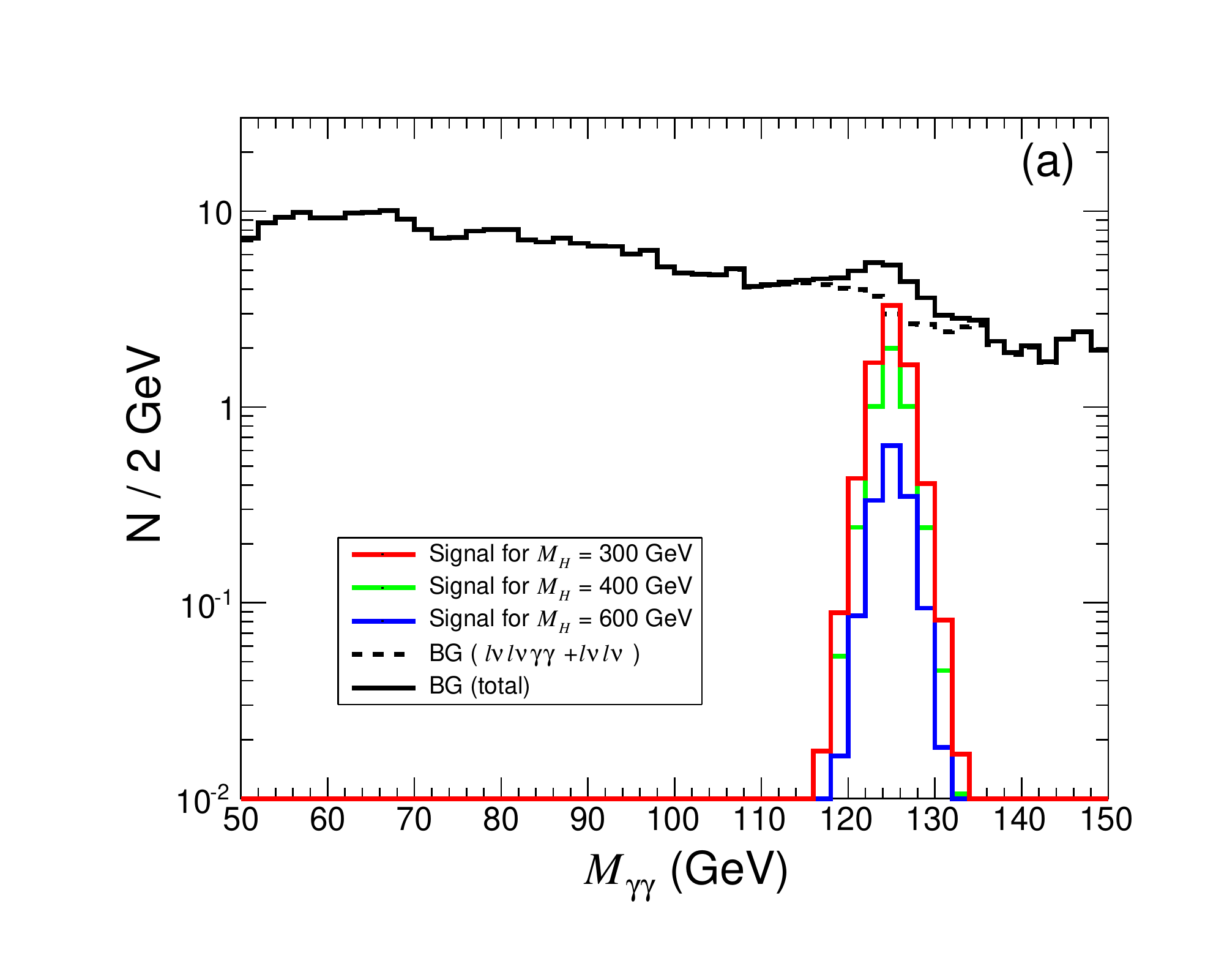}
\hspace*{-6.5mm}
\includegraphics[width=8.0cm,height=5.8cm]{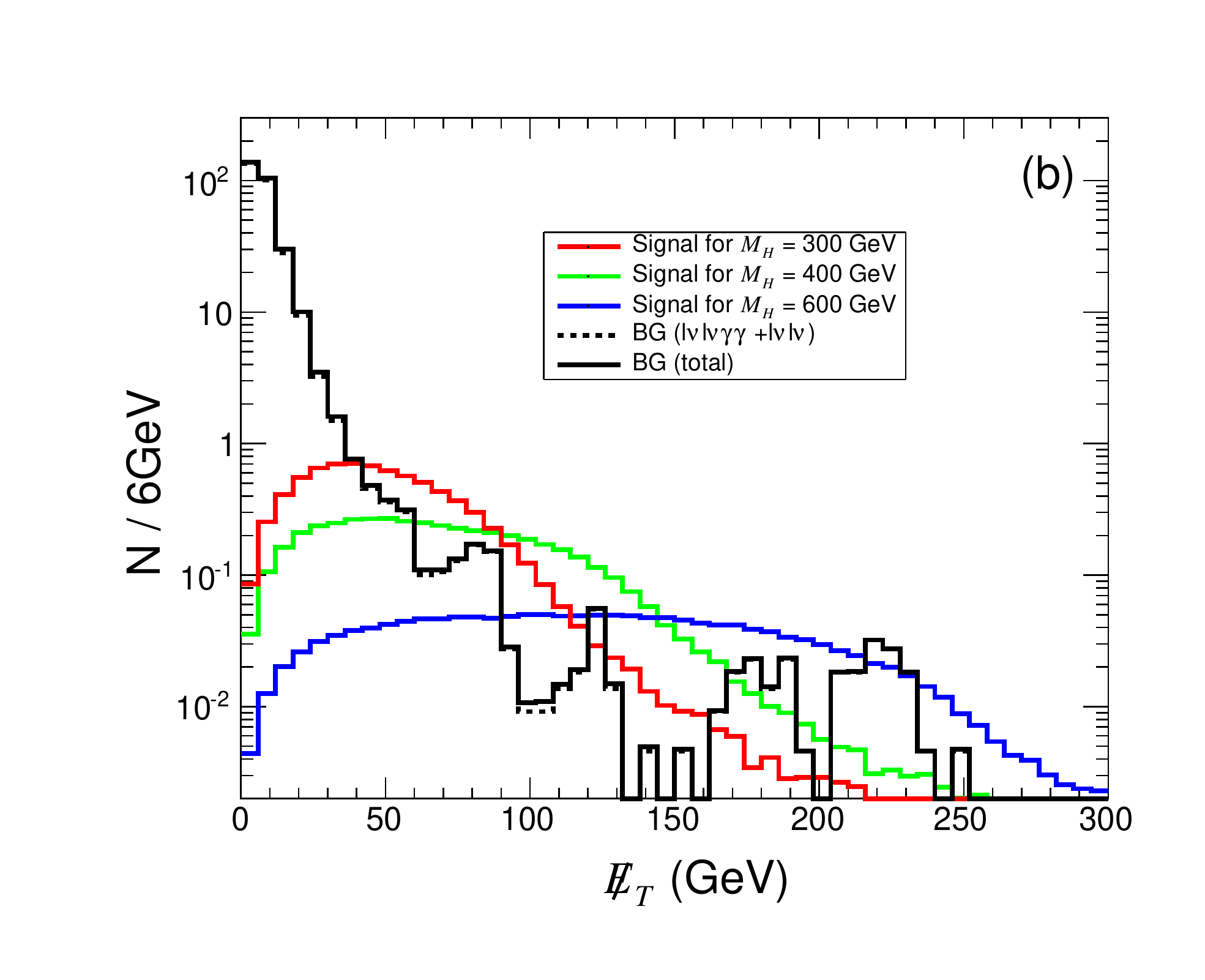}
\\[-4.5mm]
\includegraphics[width=8.0cm,height=5.8cm]{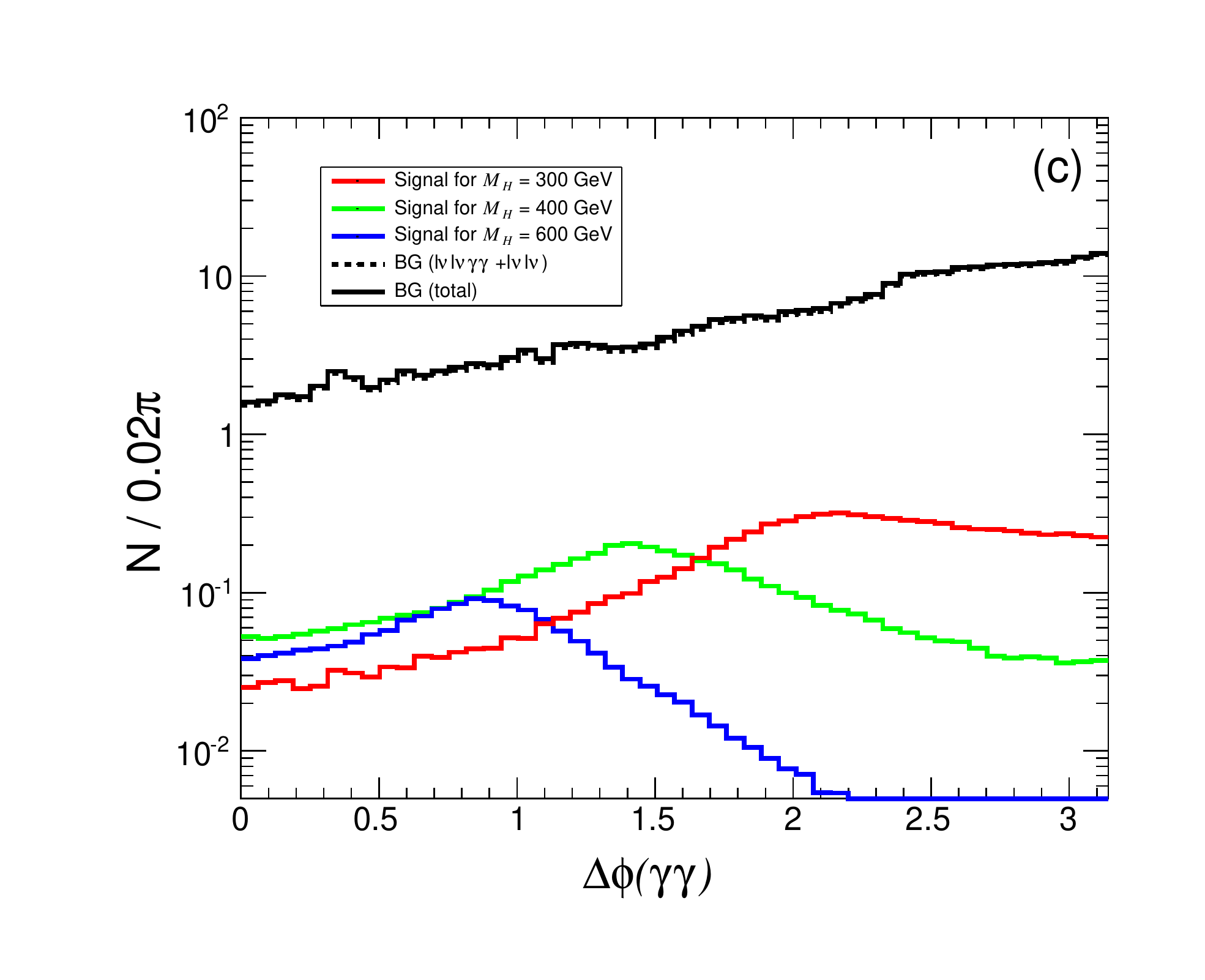}
\hspace*{-6mm}
\includegraphics[width=8.0cm,height=5.8cm]{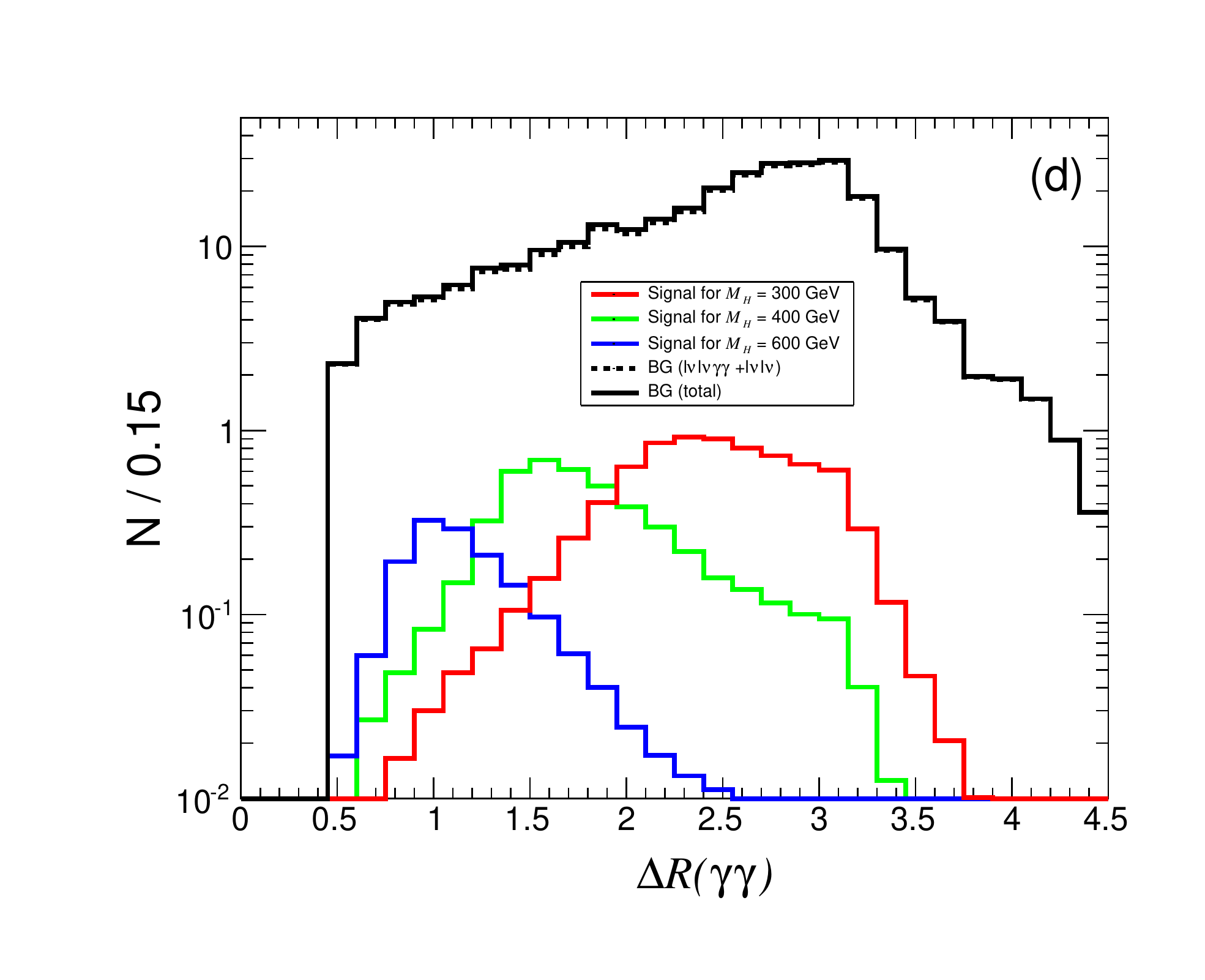}
\\[-5mm]
\includegraphics[width=8.0cm,height=5.8cm]{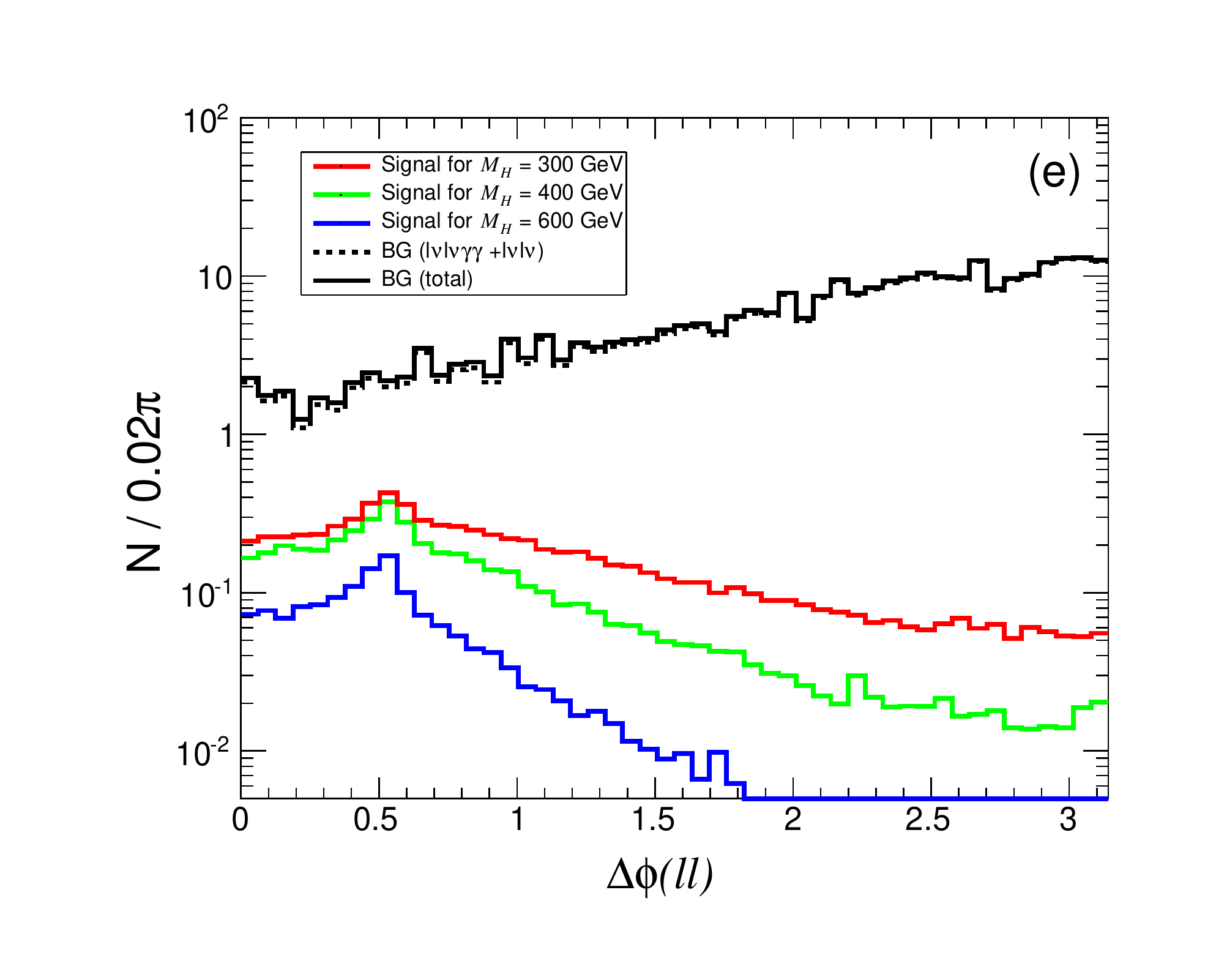}
\hspace*{-6mm}
\includegraphics[width=8.0cm,height=5.8cm]{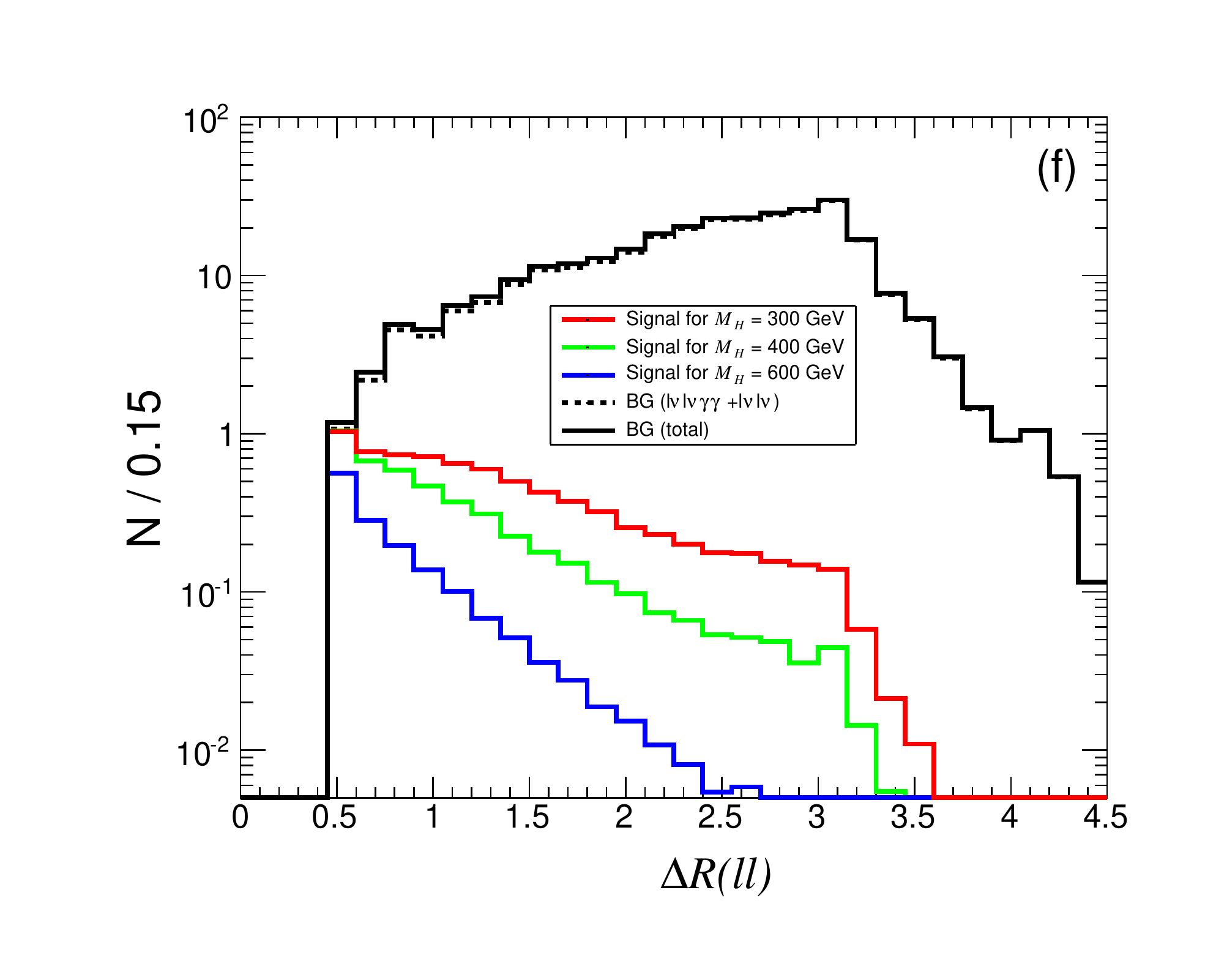}
\\[-4.5mm]
\includegraphics[width=8.0cm,height=5.8cm]{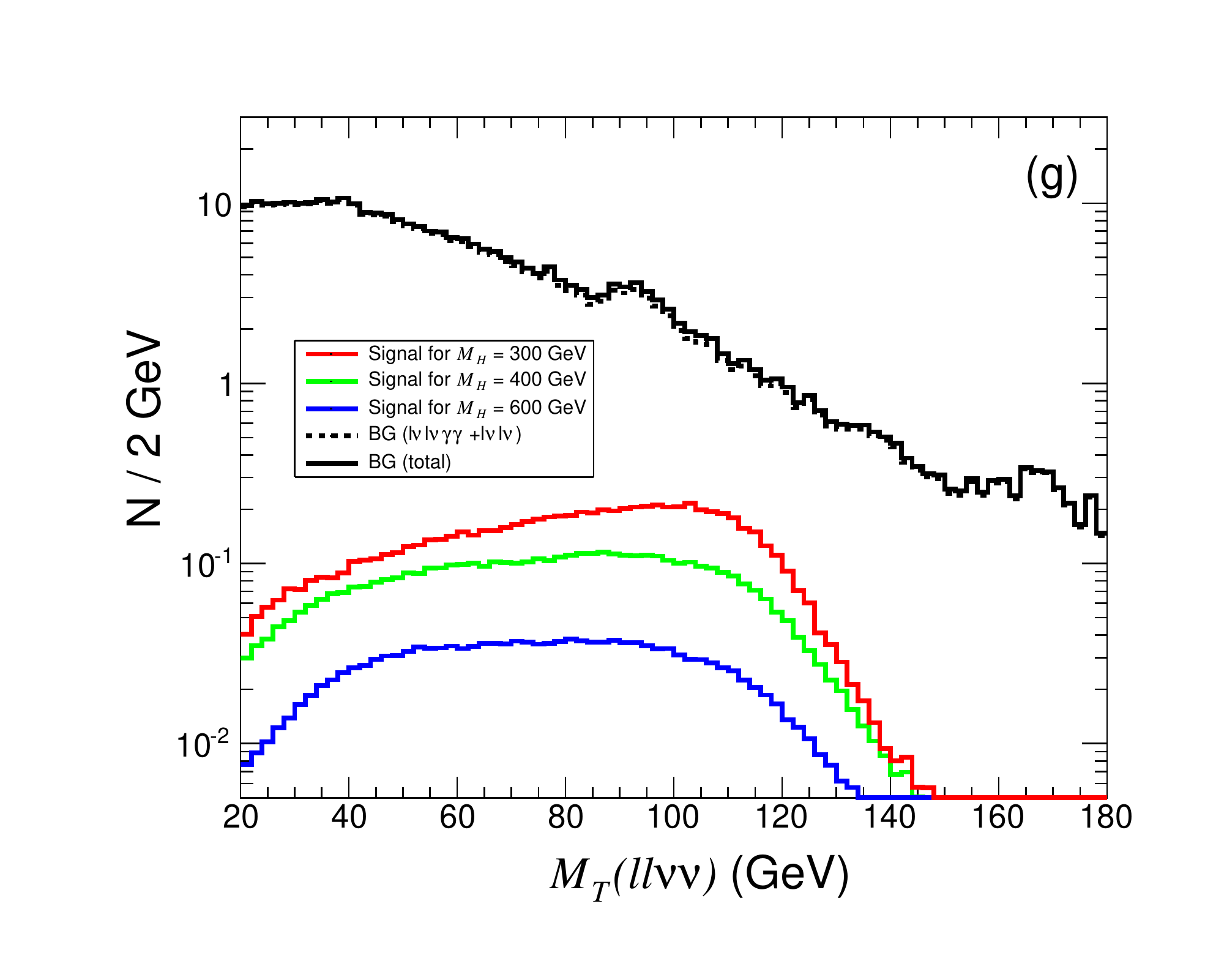}
\hspace*{-6mm}
\includegraphics[width=8.0cm,height=5.8cm]{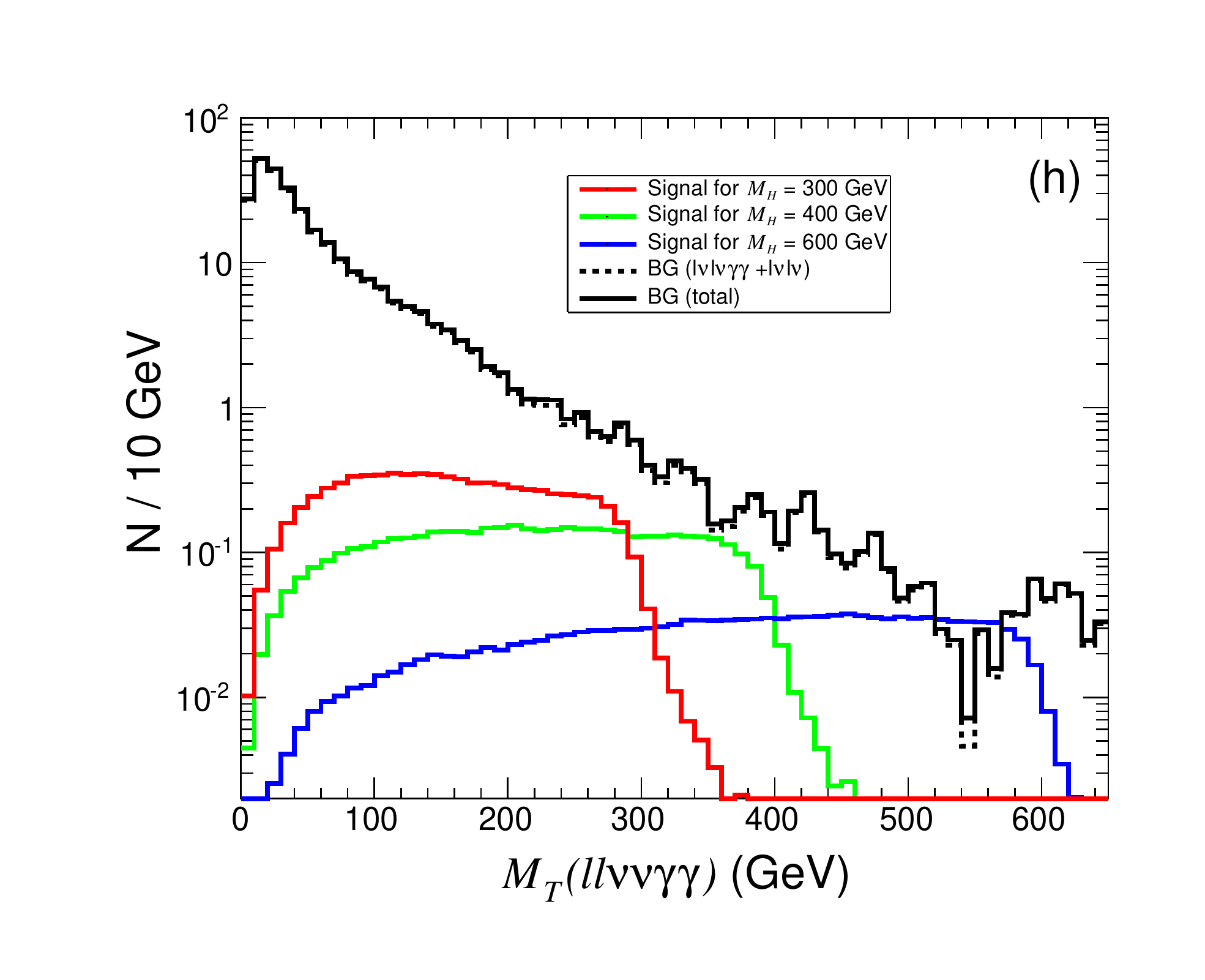}
\vspace*{-4mm}
\caption{Signal and background distributions in the pure leptonic channel
$\,hh\to WW^*\to \ell\nu\ell\nu\,$ before imposing kinematical cuts. For comparison,
we plot the signal distributions for $\,M_H^{}=(300,\,400,\,600)\,$GeV
by (red,\,green,\,blue) curves.
We present plot-(a) for $\,M_{\gamma\gamma}\,$  distribution,
plot-(b) for $E\sla_{T}^{}$ distribution,
plot-(c) for $\,\Delta\phi(\gamma\gamma)\,$ distribution,
plot-(d) for $\,\Delta R(\gamma\gamma)\,$ distribution,
plot-(e) for $\,\Delta\phi(\ell\ell)\,$ distribution,
plot-(f) for $\,\Delta R(\ell\ell)\,$ distribution,
plot-(g) for $\,M_T^{}(\ell\ell\nu\nu)\,$ distribution,
and plot-(h) for $\,M_T^{}(\ell\ell\nu\nu \gamma\gamma)\,$ distribution, respectively.
}
\label{fig:6}
\end{centering}
\vspace*{-1mm}
\end{figure}

Another potential background may arise from
the Higgs pair production $\,pp \to hh\,$ in the SM \cite{Shao:2013bz,Barger:2014taa,He:2015spf}.
Our signal process $\,pp\to H\,$ produces on-shell Higgs boson $H$ with
decays $\,H\to hh\,$,\, which has much larger rate as well as rather different kinematics from
the non-resonant di-Higgs production in the SM.
(Since our signal has on-shell $H$ production, we find that its interference
with the SM-type non-resonant $hh$ production is negligible after kinematical cuts.)
For instance, we can further suppress this SM di-Higgs contribution
by imposing a cut on the transverse mass of di-Higgs bosons.

We also consider a reducible background from the $Zh$ associate production.
The SM cross section of this process $\,pp \to Zh\,$ at the LHC is
$\,\sigma(pp \to Zh) = 0.761$\,pb \cite{Djouadi:2005gi}.
Hence, this background gives
$\,\sigma(pp \to Zh \to \ell\ell\ga\ga) = 0.175$\,fb before any cuts.
Because the $Zh$ background must have the invariant mass $\,M(\ell\ell)\,$
of final state di-leptons peaked at $\,M_Z^{} \simeq 91.2$\,GeV,\,
we can efficiently kill this background by applying a narrow cut on
$\,M(\ell\ell)\,$,\, which has little effect on the signal rate.
In the present analysis, we choose,
$\,M(\ell\ell) \in (M_{Z}^{}\!-\!5\Gamma_{Z}^{},\,M_{Z}^{}\!+\!5\Gamma_{Z}^{})$,\,
where $\Gamma_{Z}^{}\simeq 2.5\,\GeV$ is the total width of $Z$ boson.
Other reducible backgrounds come from the fake events
in which quark and/or gluon are misidentified as photons.
These backgrounds include
%
$\,\ell\nu\ell\nu q \gamma$,\,
$\ell\nu\ell\nu g \gamma$,\,
$\ell\nu\ell\nu q q$,\,
$\ell\nu\ell\nu q g$,\, and
$\,\ell\nu\ell\nu g g$\,.\,
%
For our analysis, we adopt the fake rates used by ATLAS detector \cite{Aad:2009wy},
\begin{equation}
\label{eq:mis-rate}
\epsilon_{q\rightarrow \gamma}^{} \simeq 3.6\times 10^{-4} ,
\qquad
\epsilon_{g\rightarrow \gamma}^{} \simeq 3.6\times 10^{-5} .
\end{equation}
With such small fake rates, we find that these reducible backgrounds are negligible.

In summary, with the above considerations of the SM backgrounds,
we will compute the irreducible backgrounds with final state
$\,\ell\nu \ell\nu \gamma\gamma$\,,\,
and the reducible backgrounds including the $\,\ell\ell\gamma\gamma$\,
final state, the $\,t\bar{t}h$\, associate production,
the $Zh$ associate production, and the SM di-Higgs production.

\vspace*{1.5mm}

In Fig.\,\ref{fig:6},
we present the distributions of relevant kinematical variables for the pure leptonic channel,
including both signals and backgrounds.
In this figure, we show the signal distributions at the LHC\,(14TeV) with 300\,fb$^{-1}$
integrated luminosity for $\,M_H^{}=(300,\,400,\,600)\,$GeV by (red,\,green,\,blue) curves
as well as the backgrounds (black curves). Here we have input the sample cross section
$\,\sigma(pp\!\to\!H\!\to\! hh\!\to\! WW^*\gamma\gamma)
   = (5,\,3,\,1)$\,fb for $\,M_H^{}=(300,\,400,\,600)\,$GeV, respectively.
In the following, we will analyze how to effectively suppress the SM backgrounds
by implementing proper kinematical cuts.

From Fig.\,\ref{fig:6}(a)-(b),
we first impose kinematical cuts on the diphotons invariant-mass $\,M_{\gamma\gamma}\,$
and the missing energy $E\sla_{T}$ of final state neutrinos,
\begin{equation}
\label{eq:maa-cut}
120\,\mathrm{GeV} < M_{\gamma\gamma} <130\,\mathrm{GeV} ,
\hspace*{8mm}
E\sla _{T} >20 \,\mathrm{GeV} \,.
\end{equation}
The missing energy cut can also sufficiently remove the
$\,\ell\ell\gaga\,$ background.

Then, inspecting Fig.\,\ref{fig:6}(c)-(f),
we apply the kinematical cuts on the azimuthal angle $\,\Delta\phi\,$ and
opening angle $\,\Delta R\,$ for the final state di-leptons and di-photons, respectively,
%
\beqa
\Delta\phi (\ell\ell) < 2.0\,, \qquad \Delta R(\ell\ell ) < 3.0\,,
\qquad \Delta R(\ga\ga ) < 3.8\,.
\eeqa
%
Here, from the distributions of Fig.\,\ref{fig:6}(c),
we find that the $\,\Delta\phi(\ga\ga)\,$ cut is not effective
for Higgs mass $\,M_H^{}=300\,$GeV. So we do not implement this cut.

For the transverse mass cut \cite{Beringer:1900zz},
we consider the transverse mass $\,M_T^{}\,$ for the $\ell\ell\nu\nu$ system
with two leptons and missing energy, which
should be no larger than the Higgs mass $\,M_h^{}\simeq 125$\,GeV.\,
All the final state leptons/neutrinos
are nearly massless, so the transverse energy of each final state
equals its transverse momentum  $\,E_{T,i}^{} \simeq |\vec{P}_{T,i}|$\,,\,
($i=1,2,3$),\, where $\,i=1,2\,$ denote two leptons $\,\ell_{1,2}^{}\,$ and
$\,i=3\,$ denotes the system of two neutrinos. Thus, we have
\begin{eqnarray}
M_{T}^{2} \,=\, \(E_{T,1} +E_{T,2} +E_{T,3}\)^{2} -
(\vec{P}_{T,1} +\vec{P}_{T,2} +\vec{P}_{T,3})^{2} \,\simeq
\sum_{1\leqq i<j \leqq 3} \!\! 2E_{T,i}^{}E_{T,j}^{} (1\! - \cos \phi_{ij}^{}) \,.
\end{eqnarray}
With this and inspecting Fig.\,\ref{fig:6}(g),
we implement the transverse mass cut,
\beqa
M_{T}^{}(\ell\ell\nu\nu) < 135\,\GeV  .
\eeqa
From Fig.\,\ref{fig:6}(h), we will further impose the transverse mass cut
for the full final state $\,\ell\ell\nu\nu\gamma\gamma$\,,
\begin{equation}
60 \GeV \,<\, M_T^{}(\ell\ell\nu\nu\gamma\gamma) \,<\, 320\,\mathrm{GeV} \,.
\end{equation}
The kinematical cuts for the cases of $\,M_H^{}=400$\,GeV and $\,600\,$GeV
will be discussed in Sec.\,\ref{sec:3.3}.

\vspace*{1.5mm}

We summarize the results in Table\,\ref{tab:2} for both signal and backgrounds.
For demonstration, we first input the heavier Higgs mass $\,M_H^{}=300\,$GeV,\, and set
the sample signal cross section
$\,\sigma(pp\!\to\! H\!\to\! hh\!\to\! WW^*\gamma\gamma)
   = 5\,\mathrm{fb}$\,
for the LHC\,(14\,TeV).
In Table\,\ref{tab:2}, we also show the significance of signal over backgrounds after
each set of kinematical cuts at the LHC Run-2 with 300\,fb$^{-1}$ integrated luminosity.
When the event number is small, we can use the median significance($Z_0$)
(instead of $\,S/\!\sqrt{B}$\,), as defined in following \cite{sigZ},
\begin{equation}
Z_0 = \sqrt{2\left[ (S\!+\!B)\ln\left(\frac{\,S\!+\!B\,}{B}\right) -S \!\right]\,} \, .
\end{equation}
As shown in Table\,\ref{tab:2}, after applying all kinematical cuts,
we estimate the signal significance$(Z_0^{})=5.15$\,.

\vspace*{2mm}
\subsection{{\bf Semi-leptonic Channel:}
$hh\to\! WW^*\gaga \!\to q\bar{q}'\ell\nu\,\gaga$}
\label{sec:qqlvaa}
\label{sec:3.2}
\vspace*{2mm}

\begin{figure}[t]
\vspace*{-3mm}
\begin{centering}
\hspace*{-5mm}
\includegraphics[width=8.7cm,height=7.3cm]{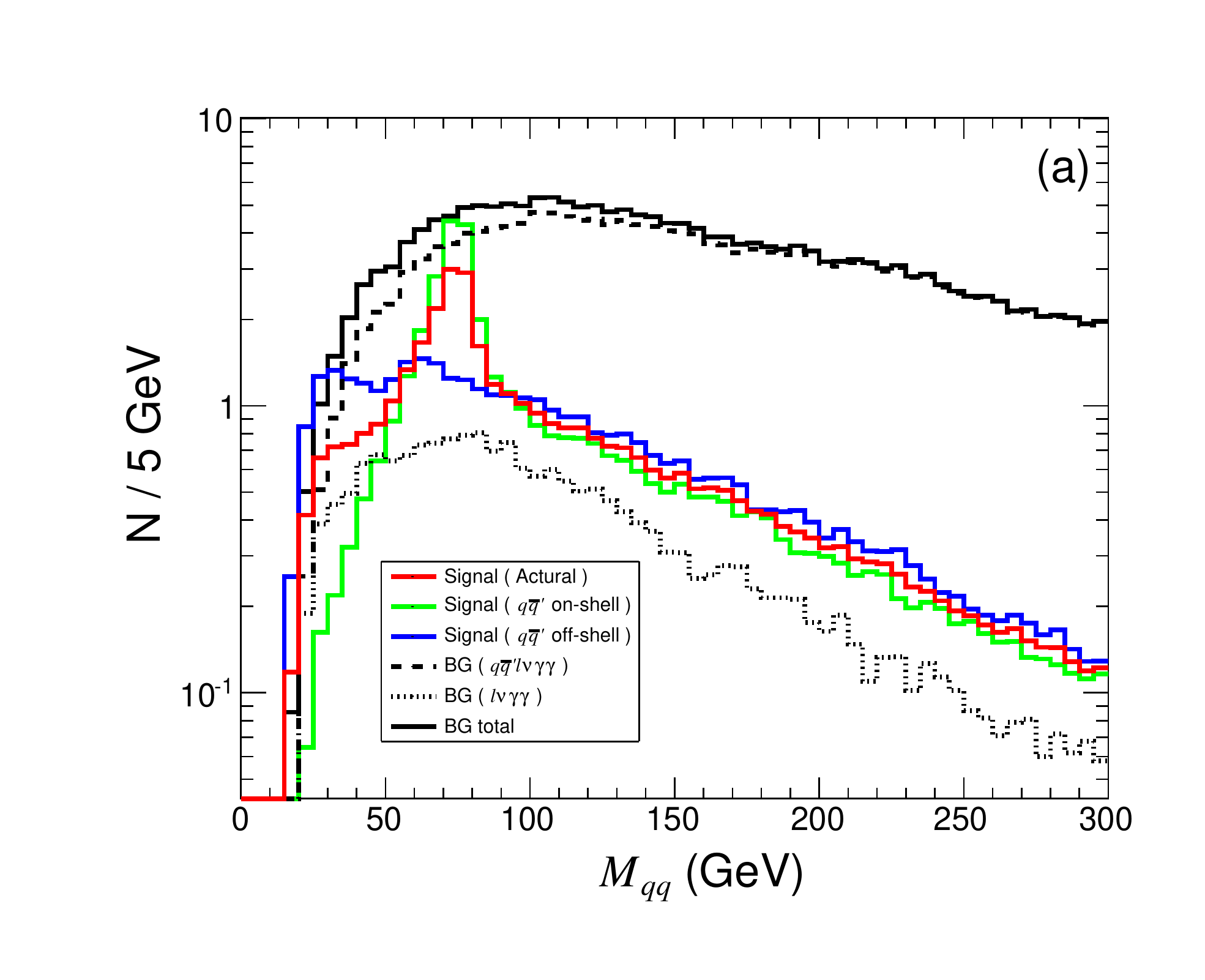}
\hspace*{-8mm}
\includegraphics[width=8.7cm,height=7.3cm]{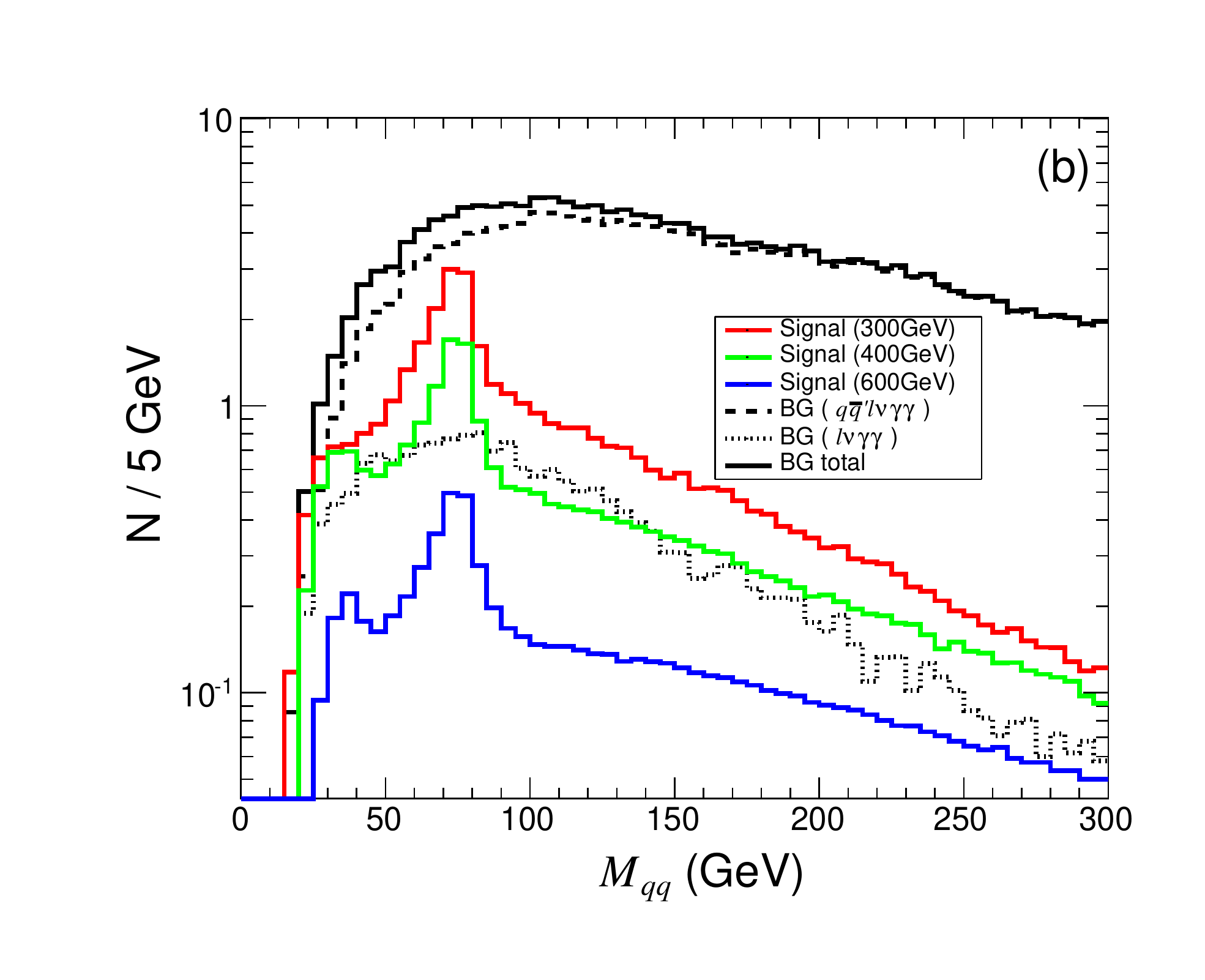}
\vspace*{-7mm}
\caption{Invariant-mass distribution of $\,M_{qq}^{}$\, for semi-leptonic decay channel
$\,WW^*\to q\bar{q}'\!\ell\nu\,$ at the LHC\,(14TeV) with 300\,fb$^{-1}$
integrated luminosity. Plot-(a) shows the mode with on-shell (off-shell) decays
$\,W\,(W^*)\to q\bar{q}'\,$  by green (blue) curve, for $\,M_H^{}=300$\,GeV.\,
The red curve corresponds to the realistic decays of $\,WW^*\to q\bar{q}'\!\ell\nu\,$.\,
Plot-(b) presents the $\,M_{qq}^{}$\, distribution for full signals of
$\,WW^*\to q\bar{q}'\!\ell\nu\,$ by (red,\,green,\,blue) curves for
$\,M_H^{}=(300,\,400,\,600)$\,GeV.
In each plot, the black solid curve gives the full backgrounds.
}
\label{fig:6-lvlvaa}
\label{fig:7}
\end{centering}
\vspace*{-3mm}
\end{figure}

The analysis of semi-leptonic channel $\,WW^*\!\to q\bar{q}'\ell\nu\,$
is similar to that of the pure leptonic mode $\,WW^*\!\to \ell\nu\ell\nu\,$.\,
But, there are nontrivial differences.
One thing is that for each decay we need to specify which decay mode is from on-shell $W$
($q\bar{q}'$ or $\ell\nu$), since these two situations have different distributions.
To illustrate this, we present the distribution of $M_{qq}^{}$
in Fig.\,\ref{fig:7}(a),
where the green (blue) curve depicts the final state $qq$ from on-shell (off-shell) $W$ decays,
and the red curve represents the actual distribution of $M_{qq}^{}$ from
$\,WW^*\!\to q\bar{q}'\ell\nu\,$.\,
Fig.\,\ref{fig:7}(a) shows that the $M_{qq}^{}$ distribution from on-shell $W$ decays
(green curve) has event rate peaked around $\,M_{qq}^{}\!=70-80$\,GeV\,,\,
while the $M_{qq}^{}$ distribution from off-shell $W$ decays (blue curve) is rather flat.

Our first step here is also to remove the pileup events, similar to Sec.\,\ref{sec:lvlvaa}.
Then, we select the final states by imposing the preliminary cuts
\begin{eqnarray}
\label{eq:pre-cuts}
n_{j}^{} \geqq 2\,, \quad
n_{\gamma}^{} \geqq 2 \,, \quad
n_{\ell}^{} =1 \,.
\end{eqnarray}
For jets we choose the leading and subleading pair, while for photons
we choose the diphoton pair whose $\,M_{\gamma\gamma}^{}\,$ is closet to $\,M_h^{} =125\,$GeV.\,
Then, we choose the basic cuts to be the same as in Eq.\,\eqref{eq:basic-cut}.

\vspace*{1.5mm}

Next, we turn to the background analysis.
The most important background for this channel comes from the SM irreducible background,
$\,pp \!\to\! q\bar{q}'\ell\nu \gamma\gamma\,$,\, whose cross section is about
\,$\sigma[q\bar{q}'\ell\nu \gamma\gamma] \simeq 31.6$\,fb.\,
Another significant reducible background is the SM process
$\,pp \to \ell\nu \gamma\gamma \,$,\, which has a cross section
\,$\sigma[\ell\nu \gamma\gamma] \simeq 143$\,fb\,.\,
But this will be mainly rejected by the jet-selections in Eq.\,\eqref{eq:pre-cuts}.
For the $\,t\bar{t}h\,$ background, we find that under $b$-veto its cross section is
$\,0.0148$\,fb, as shown in Table\,\ref{tab:2}.
Single top associated Higgs production gives another background,
$\,\sigma[pp \to th(\bar{t}h)+ X]=79.4$\,fb \cite{th},
where $\,X\,$ represents single-jet or dijets in our simulation.
We find that under $b$-veto this cross section of
$\,pp \to th(\bar{t}h)+ X\to b\ell\nu\ga\ga +X\,$
reduces to about $\,0.013$\,fb\,.\,
We also include the non-resonant di-Higgs production in the SM, which
has much smaller event rate and rather different kinematics.
Other potential SM backgrounds may include the reducible backgrounds
such as $\,qq\ell\nu gg\,$ with $\,gg\,$ misidentified as $\,\ga\ga\,$.
This is actually negligible due to the tiny $\,g\to\ga\,$ misidentification rate
shown in Eq.\,\eqref{eq:mis-rate}.

\vspace*{1.5mm}

For the kinematic cuts, we choose the $\,M_{\gamma\gamma}^{}$ cut as in Eq.\,\eqref{eq:maa-cut}.
The invariant-mass $\,M_{qq}^{}$ should match the $W$ mass. We depict the $\,M_{qq}^{}$ distribution
in Fig.\,\ref{fig:6-lvlvaa}.
Plot-(a) depicts the decay mode with on-shell (off-shell) decays
$\,W\,(W^*)\to q\bar{q}'\,$  by green (blue) curve, for $\,M_H^{}=300$\,GeV.\,
The realistic decays of $\,WW^*\to q\bar{q}'\!\ell\nu\,$ correspond to
the red curve. In plot-(b), we present the $\,M_{qq}^{}$\, distribution for full signals of
$\,WW^*\to q\bar{q}'\!\ell\nu\,$ by (red,\,green,\,blue) curves for
$\,M_H^{}=(300,\,400,\,600)$\,GeV. The black solid curve in each plot gives the full
backgrounds.  From Fig.\,\ref{fig:7}, we choose the $\,M_{qq}^{}$  cut,
\begin{equation}
M_{qq}^{} <\, 250\,\mathrm{GeV} .
\end{equation}
\begin{figure}[H]
\vspace*{-3mm}
\begin{centering}
\includegraphics[width=8.5cm,height=6.6cm]{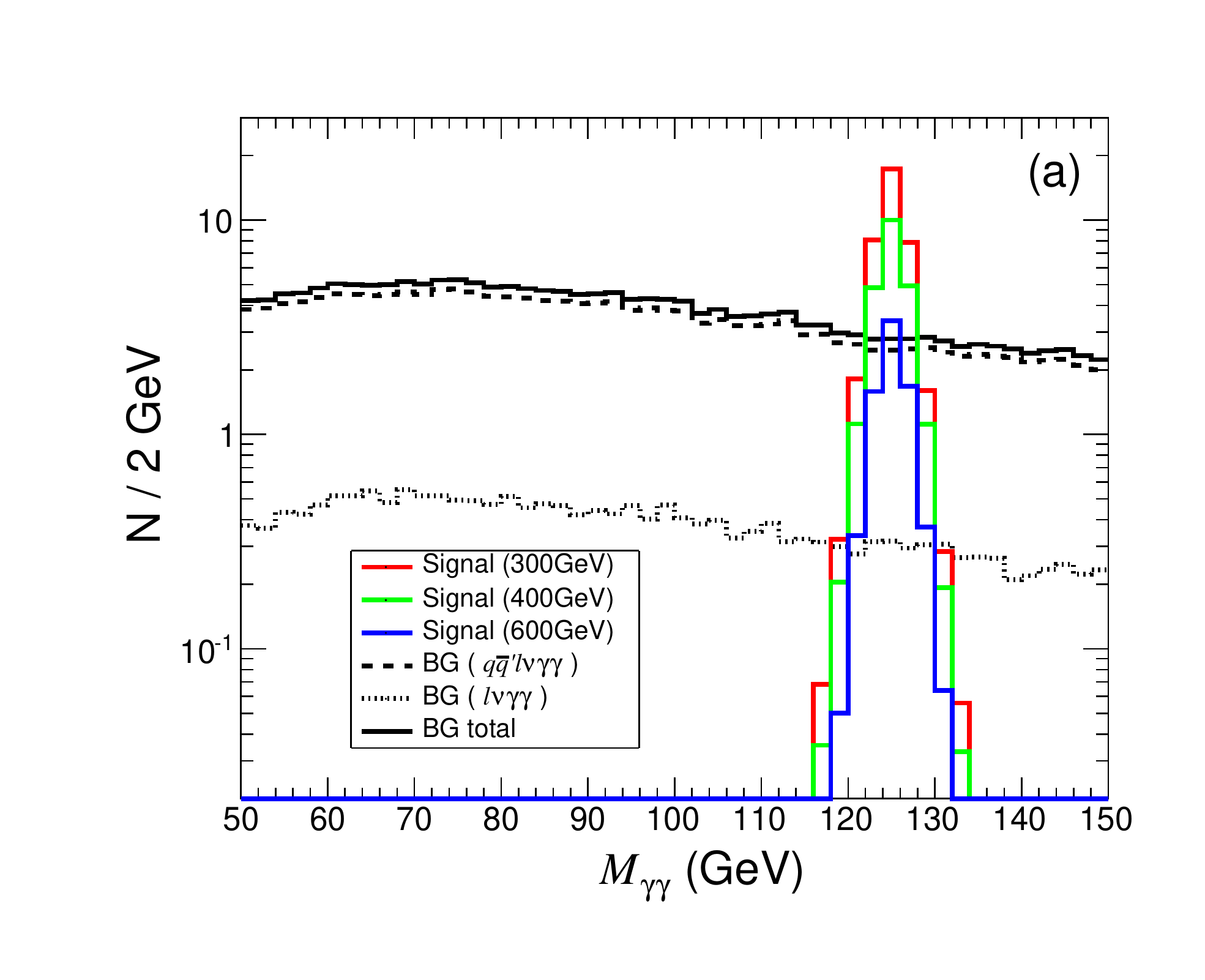}
\vspace*{-5.5mm}
\hspace*{-8mm}
\includegraphics[width=8.5cm,height=6.6cm]{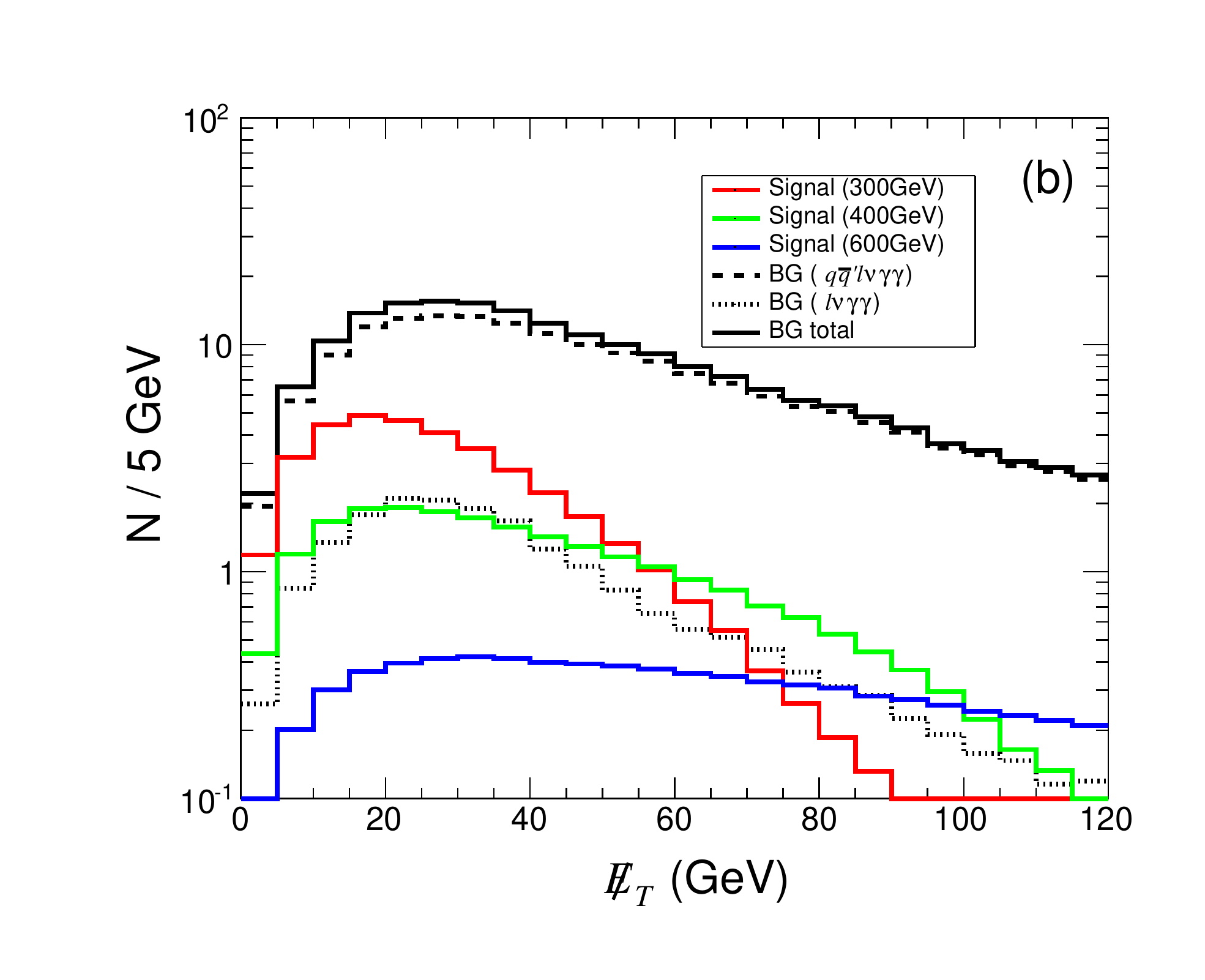}
\includegraphics[width=8.5cm,height=6.6cm]{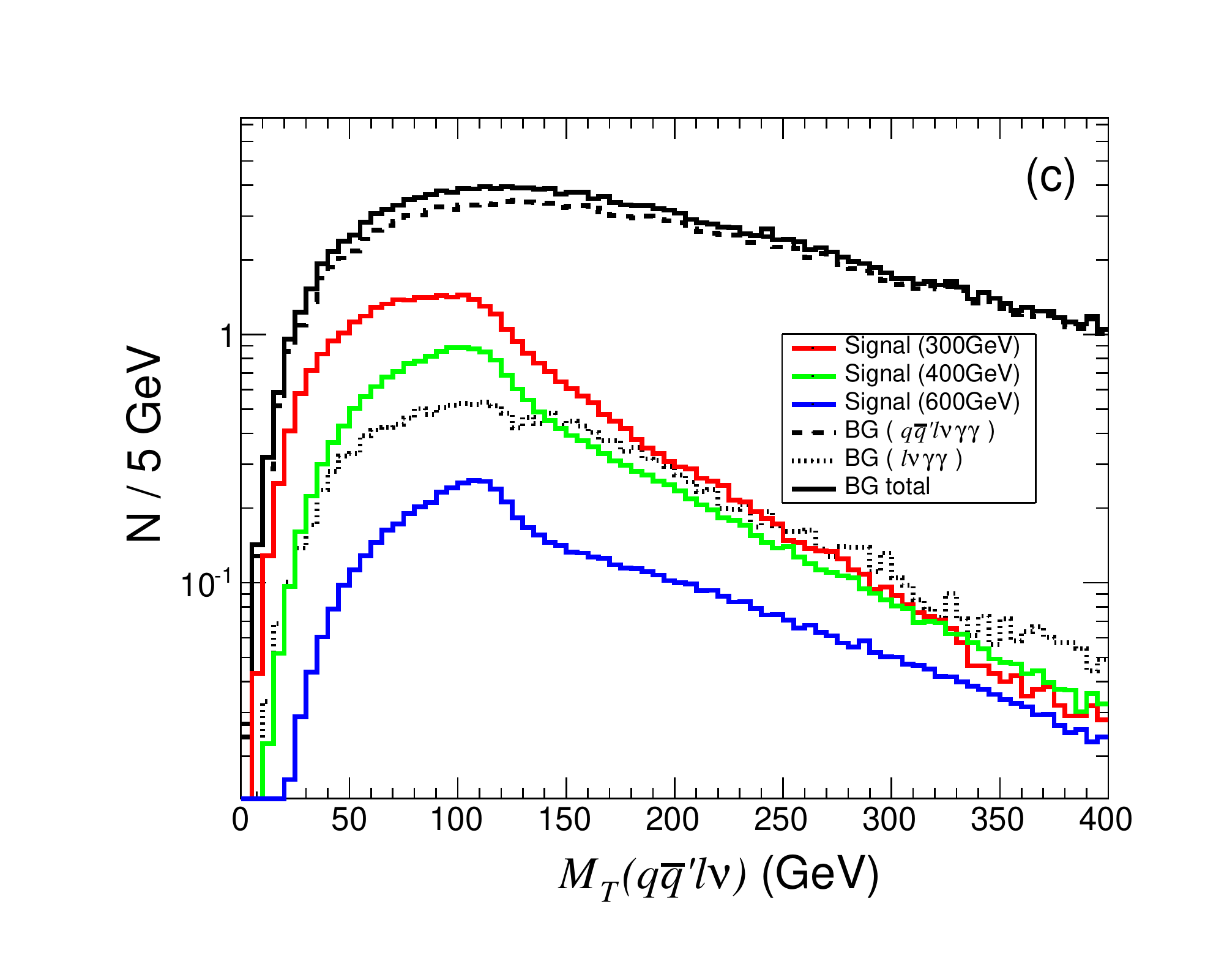}
\hspace*{-8mm}
\vspace*{-5.5mm}
\includegraphics[width=8.5cm,height=6.6cm]{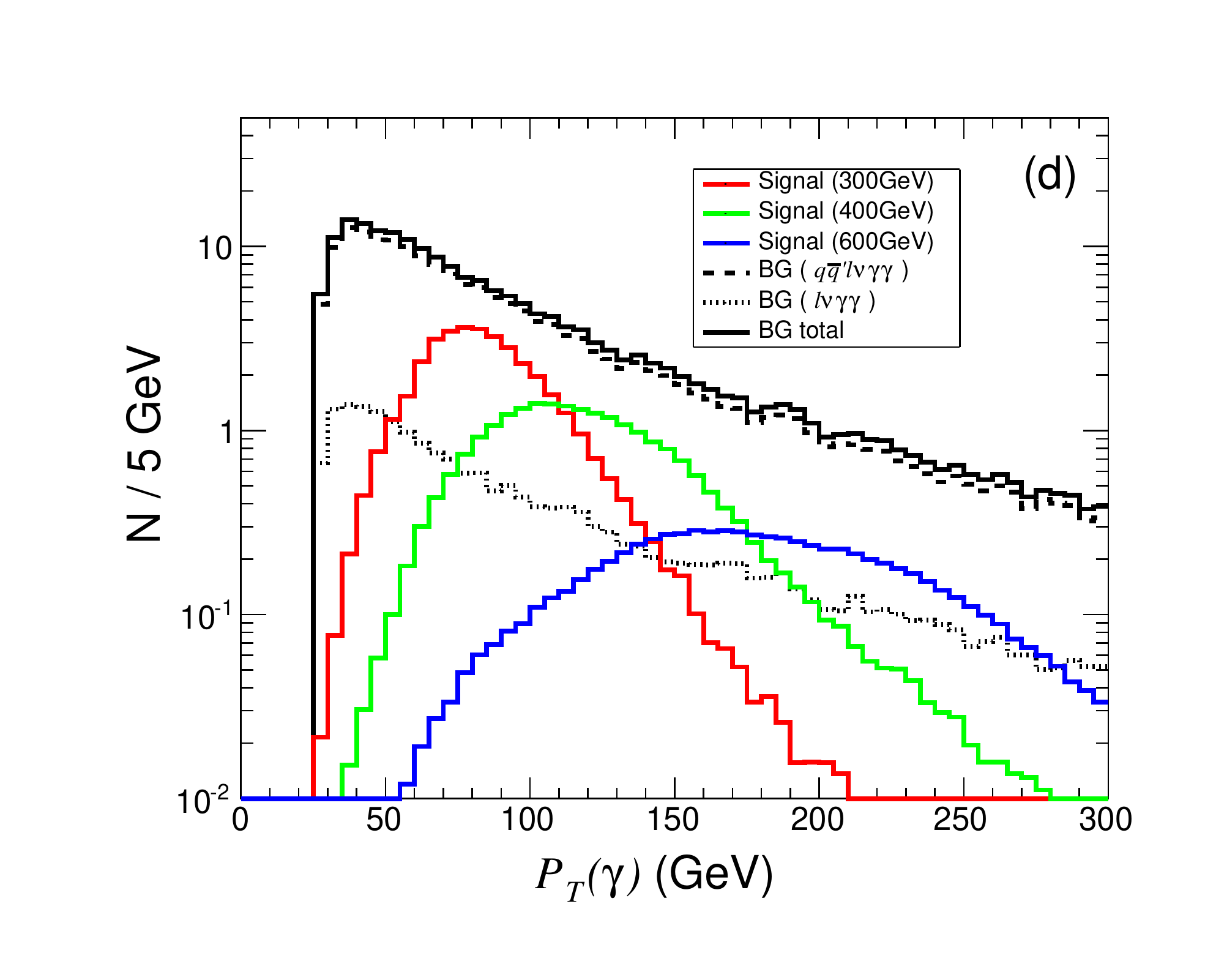}
\includegraphics[width=8.5cm,height=6.6cm]{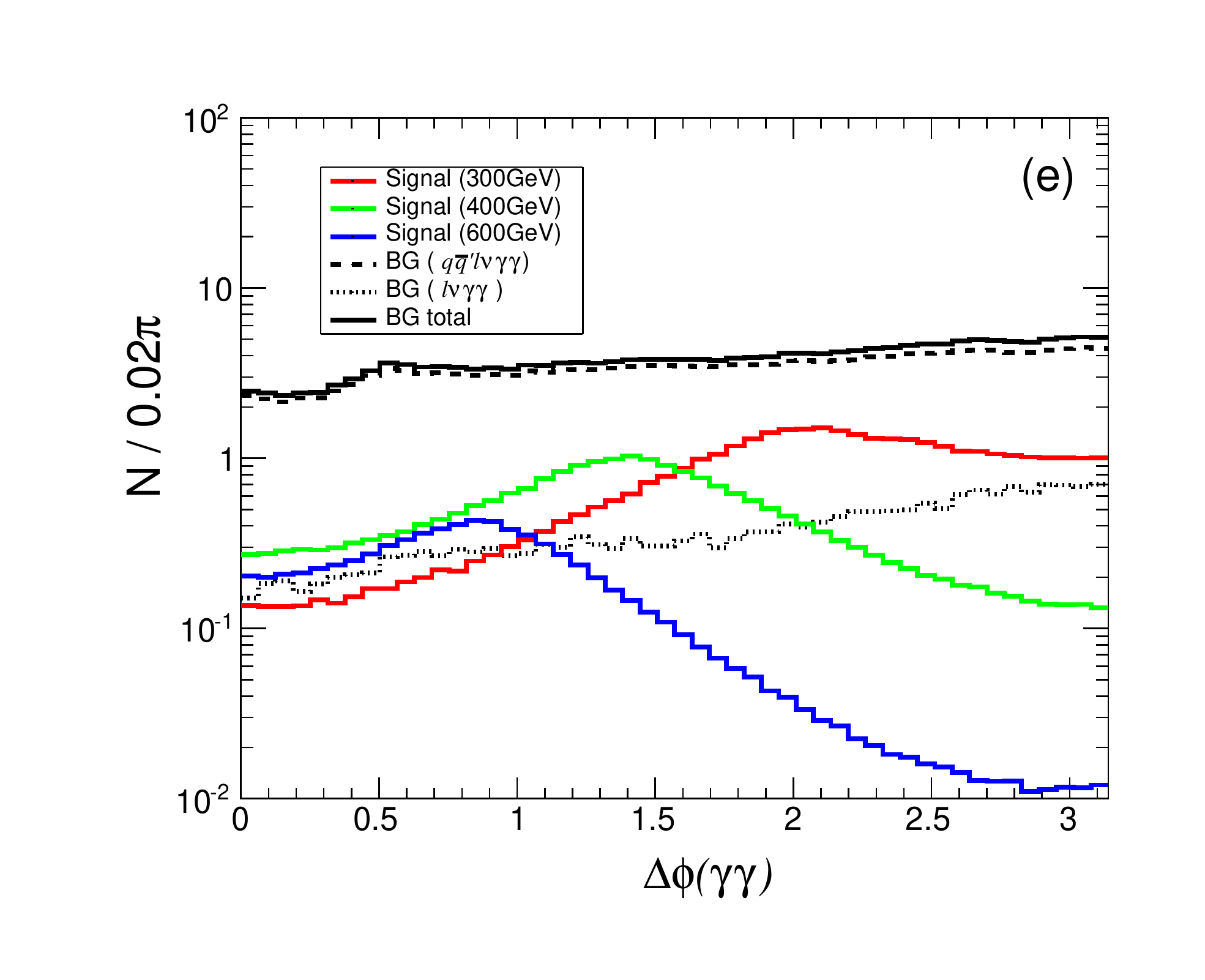}
\hspace*{-8mm}
\includegraphics[width=8.5cm,height=6.6cm]{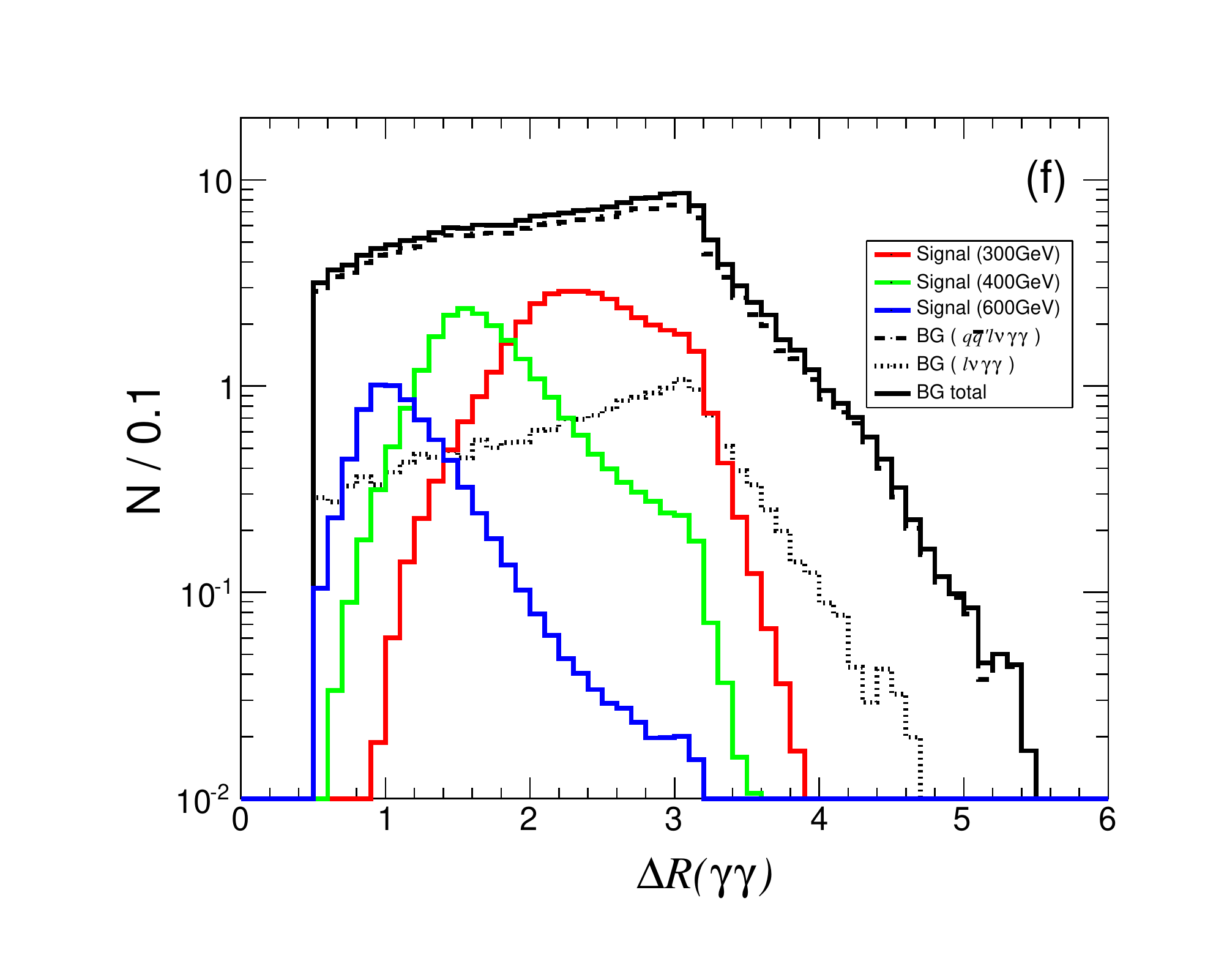}
\vspace*{-6mm}
\caption{Signal and background distributions for semi-leptonic channel
$\,hh\to WW^*\to q\bar{q}'\!\ell\nu\gamma\gamma$\,  before imposing kinematical cuts.
For comparison, we plot the signal distributions for
$\,M_H^{}=(300,\,400,\,600)\,$GeV by (red,\,green,\,blue) curves.
We present the $\,M_{\gamma\gamma}^{}$ distribution in plot-(a),
the missing $E\sla_T^{}$ distribution in plot-(b),
the $\,M_T^{}(q\bar{q}'\!\ell\nu )\,$ distribution in plot-(c),
the $P_T^{}(\gamma)$ distribution of the leading photon in plot-(d),
the $\,\Delta\phi(\ga\ga)\,$ distribution in plot-(e),
and the $\,\Delta R(\ga\ga)\,$ distribution in plot-(f), respectively.
}
\label{fig:8}
\end{centering}
\end{figure}

We present the distributions for other kinematical observables in Fig.\,\ref{fig:8},
where we have input the sample cross section
$\,\sigma(pp\!\to\!H\!\to\! hh\!\to\! WW^*\gamma\gamma)
   = (5,\,3,\,1)$\,fb\, for $\,M_H^{}=(300,\,400,\,600)\,$GeV.\,
From Fig.\,\ref{fig:8}(a)-(b), we impose cuts on the diphoton invariant-mass $\,M_{\gamma\gamma}\,$
and the missing energy $\,E\sla_T^{}$ of final state neutrinos,
\begin{equation}
\label{eq:missET}
120\,\mathrm{GeV} < M_{\gamma\gamma} <130\,\mathrm{GeV} ,
\hspace*{8mm}
10\,\GeV <  E\sla_T^{} < 80\,\mathrm{GeV} .
\end{equation}
We require $\,E\sla_T^{}>10\,$GeV to suppress certain reducible backgrounds,
as also adopted in the ATLAS analysis.
For instance, consider the background
$\,qq\gamma\gamma \!+\! j$\, with $j$ mistagged as a lepton,
where $j$ denotes a gluon or quark jet.
Since it contains no neutrino in the final state,
we can eliminate it by imposing the missing energy $\,E\sla_T^{}$ cut.
This is more like a basic cut. For the transverse momentum distribution of the leading photon
shown in Fig.\,\ref{fig:8}(d), we set the following cut,
\begin{equation}
60\,\mathrm{GeV} < \,P_T^{}(\gamma)\, < 150\,\mathrm{GeV}  .
\end{equation}

Then, we inspect the transverse mass distribution of $\,q\bar{q}'\ell\nu$\, final state,
which arises from the decay products of $\,h\!\to\! WW^*\!\!\to\! q\bar{q}'\ell\nu\,$.\,
From Fig.\,\ref{fig:8}(c), we impose the following cut,
\begin{equation}
M_T^{}(q\bar{q}'\!\ell\nu ) \,<\, 200\,\mathrm{GeV} .
\end{equation}
With Fig.\,\ref{fig:8}(e)-(f), we have also examined possible cuts on
$\,\Delta\phi(\gaga)\,$ and $\,\Delta R(\gaga)\,$ distributions. We further impose,
\beqa
1 \,<\, \Delta R(\gamma\gamma) \,<\, 3.8 \, .
\eeqa

We summarize our results in Table\,\ref{tab:2}.
Here we present the signal and background cross sections after each set of cuts.
We take an integrated luminosity of $300\,\mathrm{fb}^{-1}$ for the LHC\,(14\,TeV),
and derive the corresponding signal significance($Z_0$).
Under all cuts, we estimate the final significance of the signal detection
to be $\,7.47\,$ in the semi-leptonic channel $\,q\bar{q}'\ell\nu\gamma\gamma$\,,\,
as shown in Table\,\ref{tab:2}.

\begin{table}[t]
\centering
\caption{Signal and background cross sections of both
$\,pp\to WW^*\gaga\to \ell\nu\ell\nu\gamma\gamma$\, and
$\,pp\to WW^*\gaga\to q\bar{q}'\ell\nu\gamma\gamma$ processes
at the LHC\,(14\,TeV) after each set of cuts. The signal significance($Z_0$)
is computed for the LHC\,(14\,TeV) runs with 300\,fb$^{-1}$ integrated luminosity.
We input the heavier Higgs mass $\,M_H^{}=400\,$GeV,\, and set the sample signal cross section
$\,\sigma(pp\!\to\!H\!\to\! hh\!\to\! WW^*\gamma\gamma) = 3\,\mathrm{fb}$\,.\,
From the 3rd to 5th columns, we present the signals
and backgrounds after imposing each set of cuts.
In the pure leptonic mode, we impose the Final Cuts
$M_T^{}(\ell\ell\nu\nu)$,\, $M(\ell\ell)$,\, $M_T^{}(\ell\ell\nu\nu\gamma\gamma)$,\,
$\Delta\phi (\ell\ell)$,\, $\Delta R(\ell\ell)$,\, $\Delta\phi(\ga\ga)$,\, and
$\Delta R(\ga\ga)$.\, In the semi-leptonic mode, we add the Final Cuts
$P_T^{}(\gamma)$,\, $M_T^{}(q\bar{q}'\!\ell\nu)$,\, $\Delta\phi(\ga\ga)$,\, and $\Delta R(\ga\ga)$.
}
\vspace*{2mm}
\begin{tabular}{c||c|c|c|c}
\hline
\hline
$\,pp\to \ell\nu\ell\nu\gamma\gamma$ & Sum & Selection+Basic\,Cuts & $M_{\gamma\gamma}^{}, E\sla_{T}$ & Final Cuts
\tabularnewline
\hline
Signal (fb) &0.315 & 0.0165 & 0.0147 & 0.0107
\tabularnewline
BG[$\ell\nu\ell\nu\gaga\!+\!\ell\ell\gaga$] (fb)
& 153.3 & 0.937 & 0.00394 & 0.000169
\tabularnewline
BG$[t\bar{t}h]$ (fb)
& 0.0071 & 0.000493 & 0.000452 & 0.000051
\tabularnewline
BG$[Zh]$ (fb)
& 0.175 & 0.0331 & 0.00247 & 0.000065
\tabularnewline
BG$[hh]$ (fb)
& 0.00222 & 0.000132 & 0.000116 & 0.000074 \tabularnewline
BG[Total] (fb)
& 153.48 & 0.971 & 0.00698 & 0.000359
\tabularnewline
Significance($Z_{0}$) & 0.440 & 0.289 & 2.44 & 4.05
\tabularnewline
\hline
\hline
$\,pp\to q\bar{q}'\!\ell\nu\gamma\gamma$ &  $\sigma_{\rm total}$
&  Selection+Basic\,Cuts & $M_{\gamma\gamma}^{}$,\,$M_{qq}^{}$,\,$E\sla_T^{}$  & Final Cuts  \\
\hline
Signal (fb) & 1.32 & 0.0891 & 0.0671 & 0.0533
\\
BG[$qq\ell\nu\gamma\gamma$] (fb) & 31.59 & 0.581 & 0.0291 & 0.00672
\\
BG[$\ell\nu\gamma\gamma$] (fb) & 143.3 & 0.0642 & 0.00454 & 0.000891
\\
BG[$Wh$] (fb) & 0.42 & 0.00509 & 0.00335 & 0.00139\\
BG[$WWh$] (fb) & 0.0023 & 0.000210 & 0.000127 & 0.000057\\
BG[$t\bar{t}h$] (fb) &  0.0148 & 0.00163 & 0.00111 & 0.000441
\\
BG[$hh$] (fb) &   0.00462 & 0.000291 & 0.000197 & 0.000155
\\
BG[$th$] (fb) &  0.0129 & 0.000479 & 0.000247 & 0.000104
\\
BG[Total] (fb) & 175.35 & 0.653 & 0.0386 & 0.0098
\\
Significance($Z_{0}$) & 1.72 & 1.87 & 4.86  & 6.22
\\
\hline
\hline
\end{tabular}
\label{tab:3-400GeV}
\label{tab:3}
\end{table}

\vspace*{3mm}
\subsection{\bf Analyses of Heavier Higgs Boson with 400\,GeV and 600\,GeV Masses}
\label{sec:400GeV}
\label{sec:3.3}
\vspace*{2mm}

For signal and background analyses in Sec.\,\ref{sec:lvlvaa}--\ref{sec:qqlvaa},
we have set the mass of heavier Higgs boson $\,M_H^{}=300\,$GeV for demonstration.
In this subsection, we turn to the analyses for other sample inputs of Higgs mass,
$\,M_H^{}=400\,\GeV$\, and $\,M_H^{}=600\,\GeV$.\,
We demonstrate how the analysis and results may vary as the Higgs mass increases.
These are parallel to what we have done in Sec.\,\ref{sec:lvlvaa}--\ref{sec:qqlvaa}.

For the heavier Higgs boson with mass $\,M_H^{}=400$\,GeV,\,
from the distributions in Fig.\,\ref{fig:6},
we choose the following kinematical cuts for
the pure leptonic channel,
%
\beqa
& & 120\,\mathrm{GeV} < M_{\gamma\gamma}^{} < 130\,\mathrm{GeV},
\hspace*{5mm}
\Delta\phi(\ga\ga) < 2.5 \,,
\hspace*{5mm}
\Delta R(\gamma\gamma) < 2.5 \,,
\nonumber\\
&&  E\sla_{T}^{} >20\,\mathrm{GeV},
\hspace*{5mm}
M_{T}^{}(\ell\ell\nu\nu) < 135\,\GeV  ,
\hspace*{5mm}
75\GeV < M_T^{}(\ell\nu\ell\nu\gamma\gamma) <420\,\mathrm{GeV},
\hspace*{7mm}
\label{eq:400GeV-2L-cut}
\\
\label{eq:400GeV-cut-b}
&&
\Delta \phi(\ell\ell) < 2.0\,,
\hspace*{5mm}
\Delta	R (\ell\ell) < 2.2 \,,
\hspace*{5mm}
M(\ell\ell) ~\setminus\hspace*{-2.7mm}\in
(M_{Z}^{}\!-\!5\Gamma_{Z}^{},M_{Z}^{}\!+\!5\Gamma_{Z}^{})\,.
\nonumber
\eeqa
%

Comparing with the previous case of $\,M_H^{} =300$\,GeV,\,
we find that the distributions
$\,\Delta\phi (\ell\ell)$,\, $\,\Delta R(\ell\ell)$,\,
$\,\Delta\phi(\gaga)$,\, and $\,\Delta R(\gaga)\,$
damp faster in the larger $\,\Delta\phi\,$ and $\,\Delta R$\, regions,
as shown in Fig.\,\ref{fig:6}.
This is because the di-Higgs bosons are more boosted in the
$\,H\to hh\,$ decays with heavier mass $\,M_H^{} =400$\,GeV\,.\,
We present the cut efficiency for the case of $\,M_H^{} =400\,$GeV\,
in Table\,\ref{tab:3}, where we set a sample signal cross section
$\,\sigma(pp\!\to\! H\!\to\! hh\!\to\! WW^*\gamma\gamma)
   = 3\,\mathrm{fb}$\,.\,
In this case, we derive a signal significance$(Z_{0}^{}) =4.05$\,
after all the kinematical cuts.
We also note from Fig.\,\ref{fig:4}(a)-(b) that in 2HDM-I the cross section
$\,\sigma(pp\!\to\! HX)\times\text{Br}(H\!\to\! hh\!\to\! WW^*\gamma\gamma)\,$
can reach up to 30\,fb\, for $\,M_H^{} =400\,$GeV\,,\,
while in 2HDM-II this cross section is below about
2\,fb at $\,M_H^{} =400\,$GeV\,.\, Hence, the significance for probing 2HDM-II with
$\,M_H^{} =400\,$GeV will be rescaled accordingly, as we will do in Sec.\,\ref{sec:4}.

\vspace*{1.5mm}

Then, we further analyze semi-leptonic channels for detecting the heavier Higgs boson $\,H\,$
with mass $\,M_H^{}=400$\,GeV\,.\,
The corresponding signal and background distributions are presented in Fig.\,\ref{fig:8}.
Inspecting these distributions, we choose the following kinematical cuts,
%
\beqa
& & 120\,\mathrm{GeV} < M_{\gamma\gamma}^{} < 130\,\mathrm{GeV}, \quad
M_{qq}^{} < 250\,\GeV ,   \quad
\nonumber \\
& & 60\,\mathrm{GeV} < P_T^{}(\gamma) < 250\,\mathrm{GeV} ,  \quad
M_T^{}(q\bar{q}'\!\ell\nu) < 250\,\mathrm{GeV} , \quad
E\sla_T^{} > 10\,\GeV ,  \quad
\label{eq:400GeV-cut}
\\
& & \Delta\phi (\gamma\gamma) < 2.3\,, \qquad
    0.75 < \Delta R(\gamma\gamma) < 2.2 \,.
\nonumber
\eeqa
%
We summarize cut efficiency of the final state $\,qq\ell\nu\gamma\gamma\,$  for
$\,M_H^{}=400$\,GeV in Table\,\ref{tab:3}.
We derive a significance  $\,Z_{0}^{}=6.22$\, after all kinematical cuts.

\begin{table}
\centering
\caption{Signal and background cross sections of both
$\,pp\to WW^*\gaga\to \ell\nu\ell\nu\gamma\gamma$\, and
$\,pp\to WW^*\gaga\to q\bar{q}'\ell\nu\gamma\gamma$\, processes
at the LHC\,(14\,TeV) after each set of cuts. The signal significance($Z_0$)
is computed for the LHC\,(14\,TeV) with an integrated luminosity of 3\,ab$^{-1}$.\,
We input the heavier Higgs mass $\,M_H^{}=600\,$GeV,\,  and set the sample signal cross section
$\,\sigma(pp\!\to\!H\!\to\! hh\!\to\! WW^*\gamma\gamma) = 1\,\mathrm{fb}$.
From the 3rd to 5th columns, we present the signals
and backgrounds after imposing each set of cuts.
In the pure leptonic mode, we impose the Final Cuts $M_T^{}(\ell\ell\nu\nu)$,\, $M(\ell\ell)$,\,  $M_T^{}(\ell\ell\nu\nu\gamma\gamma)$,\, $\Delta\phi (\ell\ell)$,\,
$\Delta R(\ell\ell)$,\, $\Delta\phi(\ga\ga)$,\, and $\Delta R(\ga\ga)$.\,
In the semi-leptonic mode, we add the Final Cuts $P_T^{}(\gamma)$,\,
$M_T^{}(q\bar{q}'\!\ell\nu)$,\,  $\Delta\phi(\ga\ga)$,\, and $\Delta R(\ga\ga)$.
}
\vspace*{2mm}
\begin{tabular}{c|c|c|c|c}
\hline
\hline
$\,pp\to \ell\nu\ell\nu\gamma\gamma$ & Sum & Selection+Basic\,Cuts & $M_{\gamma\gamma}^{}, E\sla_{T}$ & Final Cuts\\
\hline
Signal\ (fb)& 0.105 & 0.00578 & 0.00540 & 0.00451\\
BG[$\ell\nu\ell\nu\gaga\!+\!\ell\ell\gaga$] (fb)
& 153.3 & 0.937 & 0.00348 & 0.000092
\\
BG$[t\bar{t}h]$ (fb)
&  0.0071 & 0.000493 & 0.000452 & 0.000028
\tabularnewline
BG$[Zh]$ (fb)
&  0.175 & 0.0331 & 0.00138 & 0.000029
\tabularnewline
BG$[hh]$ (fb)
& 0.00222 & 0.000132 & 0.000117 & 0.000070
\tabularnewline
BG[Total] (fb)
& 153.48 & 0.971 & 0.00543 & 0.000219
\tabularnewline
Significance$(Z_{0})$ & 0.464 & 0.321 & 3.53 & 7.76 \\
\hline
\hline
 $\,pp\to q\bar{q}'\ell\nu\gamma\gamma$ &  $\sigma_{\rm total}$
 &  Selection+Basic\,Cuts & $M_{\gamma\gamma}^{}$,\,$M_{qq}^{}$,\,$E\sla_T^{}$
 & Final Cuts  \\
\hline
Signal	\ (fb)& 0.44 & 0.0260 & 0.0163 & 0.0148
\\
BG[$qq\ell\nu\gamma\gamma$] (fb) & 31.59 & 0.581 & 0.00950 & 0.00241\\
BG[$\ell\nu\gamma\gamma$] (fb) & 143.3 & 0.0642 & 0.00176 & 0.000395\\
BG[$Wh$] (fb) & 0.42 & 0.00509 & 0.00119 & 0.000696\\
BG[$WWh$] (fb) & 0.0023 & 0.000210 & 0.000035 & 0.000035\\
BG[$t\bar{t}h$] (fb) & 0.0148 & 0.00163 & 0.000402 & 0.000237
\\
BG[$hh$] (fb) &  0.00462 & 0.000291 & 0.000120 & 0.000087
\\
BG[$th$] (fb) & 0.0129 & 0.000479 & 0.000094 & 0.000058
\\
BG[Total] (fb) & 175.35 & 0.653 & 0.0131 & 0.00392
\\
Significance($Z_{0}$) & 1.82 & 1.75 & 6.70 & 9.29
\\
\hline
\hline
\end{tabular}
\label{tab:4-600GeV}
\label{tab:4}
\end{table}

\vspace*{2mm}

Next, for the heavier Higgs $\,H\,$ with mass $\,M_H^{}=600$\,GeV\,,\,
the distributions of pure leptonic mode are shown in Fig.\,\ref{fig:6}.
From these, we set up the following kinematical cuts,
%
\beqa
& & 120\,\mathrm{GeV} < M_{\gamma\gamma} < 130\,\mathrm{GeV},
\hspace*{5mm}
E\sla_{T} >25\,\mathrm{GeV} ,
\hspace*{5mm}
\nonumber\\
&& M_{T}(\ell\ell\nu\nu) < 135\GeV  ,
\hspace*{5mm}
75\,\GeV < M_T^{}(\ell\nu\ell\nu\gamma\gamma) < 620\,\mathrm{GeV},
\hspace*{7mm}
\label{eq:600GeV-2L-cut}
\\
&& \Delta \phi(\ell\ell) < 1.5\, ,
\hspace*{6mm}
\Delta	R (\ell\ell) < 1.8 \,,
\hspace*{5mm}
M(\ell\ell) ~\setminus\hspace*{-2.7mm}\in
(M_{Z}^{}\!-\!5\Gamma_{Z}^{},M_{Z}^{}\!+\!5\Gamma_{Z}^{})\,,
\nonumber\\
&& \Delta\phi(\ga\ga) < 1.8 \,,
\hspace*{5mm}
\Delta R(\gamma\gamma) < 2.5  \,.
\nonumber
\eeqa
%
The cut efficiency for $\,M_{H}^{}=600$\,GeV\,
is summarized in Table\,\ref{tab:4}.

\vspace*{1.5mm}

For the semi-leptonic final state $\,q\bar{q}'\ell\nu\gamma\gamma$\,
with $\,M_H^{}=600$\,GeV\,,\,  we choose the kinematical cuts,
%
\beqa
& &  120\,\mathrm{GeV} < M_{\gamma\gamma}^{}\! < 130\,\mathrm{GeV} ,
\quad
 M_{qq} < 250\,\GeV ,   \quad
\nonumber \\
& & P_T^{}(\gamma) > 120 \,\mathrm{GeV},  \quad
M_T^{}(q\bar{q}'\!\ell\nu) < 350\,\mathrm{GeV} , \quad
E\sla_T^{} > 10\,\GeV ,  \quad
\label{eq:400GeV-cut}
\\
& & \Delta\phi (\gamma\gamma) < 1.6 \,, \qquad
    \Delta R(\gamma\gamma) < 1.7 \,.
\nonumber
\eeqa
%
With these, we summarize the cut efficiency of $\,q\bar{q}'\ell\nu\gamma\gamma$\, final state
for $\,M_H^{}=600\,$GeV\, in Table\,\ref{tab:4}.
Since the typical production cross section with $\,M_H^{}=600\,\GeV$\,
becomes significantly smaller over the parameter space, we take a sample input
$\,\sigma(pp\!\to\! HX)\!\times\!\text{Br}(H\!\to\! hh\!\to\! WW^*\gamma\gamma) = 1\,$fb\,,\,
and consider an integrated luminosity of \,3\,ab$^{-1}$  at the LHC\,(14\,TeV).
Hence, from Table\,\ref{tab:4}, we can estimate the significance
$\,Z_0^{}=7.76\,$ and $\,Z_0^{}=9.29\,$  for channels
$\,WW^*\gaga\to \ell\nu\ell\nu\gamma\gamma\,$ and
$\,WW^*\gaga\to q\bar{q}'\ell\nu\gamma\gamma\,$,\, respectively.
Besides, from Fig.\,\ref{fig:4}(a)-(b) we see that for $\,M_H^{} =600\,$GeV,\,
the cross section
$\,\sigma(pp\!\to\! HX)\!\times\!\text{Br}(H\!\to\! hh\!\to\! WW^*\gamma\gamma)\,$
in 2HDM-I can reach up to 3\,fb, while this cross section in 2HDM-II is below about 0.2\,fb.
Thus, the significance for probing the 2HDM-II with
$\,M_H^{} =600\,$GeV will be rescaled accordingly.
In the following Sec.\,\ref{sec:4}, we will give a general analysis of
the significance by scanning the parameter space of 2HDM-I and 2HDM-II
without assuming a sample cross section.

\vspace*{1.5mm}

In the above analyses of Table\,\ref{tab:2}--\ref{tab:4},
we have taken the sample cross sections,
$\,\sigma(pp\!\to\! H\!\to\! hh\!\to\! WW^*\gamma\gamma)
   = (5,\,3,\,1)\,$fb,
and an integrated luminosity
$\,{\cal L}=(300,\,300,\,3000)$\,fb$^{-1}$\,
for $\,M_H^{}=(300,\,400,\,600)\,$GeV\,.\, We have derived the significance of detecting
$\,H\,$ in each case. Thus, we may estimate the combined significance($Z_0$) by including both
pure leptonic and semi-leptonic decay channels,
\beqs
\beqa
\label{eq:Z0-comb}
Z_0^{}(\mathrm{combined}) &=&
\sqrt{Z_0^2(\ell\nu\ell\nu\gamma\gamma) +Z_0^2(q\bar{q}'\!\ell\nu\gamma\gamma)\,}
\nonumber\\
&\simeq& (9.06,\,7.41,\,12.1)  \,,~~~~~~~~
\text{for}~~{\cal L}=(300,\,300,\,3000)\,\text{fb}^{-1};
\\
&\simeq& (7.40,\,6.05,\,6.99)  \,,~~~~~~~~
\text{for}~~{\cal L}=(200,\,200,\,1000)\,\text{fb}^{-1};
\eeqa
\eeqs
which corresponds to $\,M_H^{}=(300,\,400,\,600)$\,GeV, respectively.

\vspace*{2mm}
\section{Probing 2HDM Parameter Space at the LHC}
\label{sec:4}
\vspace*{1.5mm}

In this section, we study the probe of 2HDM parameter space
by using the LHC Run-2 detection of the heavier Higgs state $H^0$ via
$\,pp(gg) \!\to\! H \!\to\! hh \!\to\! WW^*\gaga$\, (Sec.\,\ref{sec:3}),
as well as the current global fit for the lighter Higgs
boson $h^0$(125GeV) at the LHC Run-1.
For the present analysis, we will convert the collider sensitivity
(Sec.\,\ref{sec:3}) into the constraints on the parameter space of
2HDM-I and 2HDM-II.  As we showed in Fig.\,\ref{fig:3}(c)-(d) and explained in
the last paragraph of Sec.\,\ref{sec:2},
the inclusive Higgs production cross section $\,\sigma(pp\to HX)\,$
is always dominated by the gluon fusion channel $\,gg\to H\,$ in the
small $\tanb$ region, while other $b$-related channels are negligible.
(For 2HDM-I, this feature actually holds for full range of $\,\tanb\geqq 1$.\,)
Hence, for the present analysis, we will use Higgs production via gluon fusion
$\,pp(gg)\to H\to hh\to WW^*\gaga$\,
(Sec.\,\ref{sec:3}) to probe the 2HDM parameter space.

\begin{figure}[t]
\begin{centering}
\includegraphics[width=8.0cm,height=6.4cm]{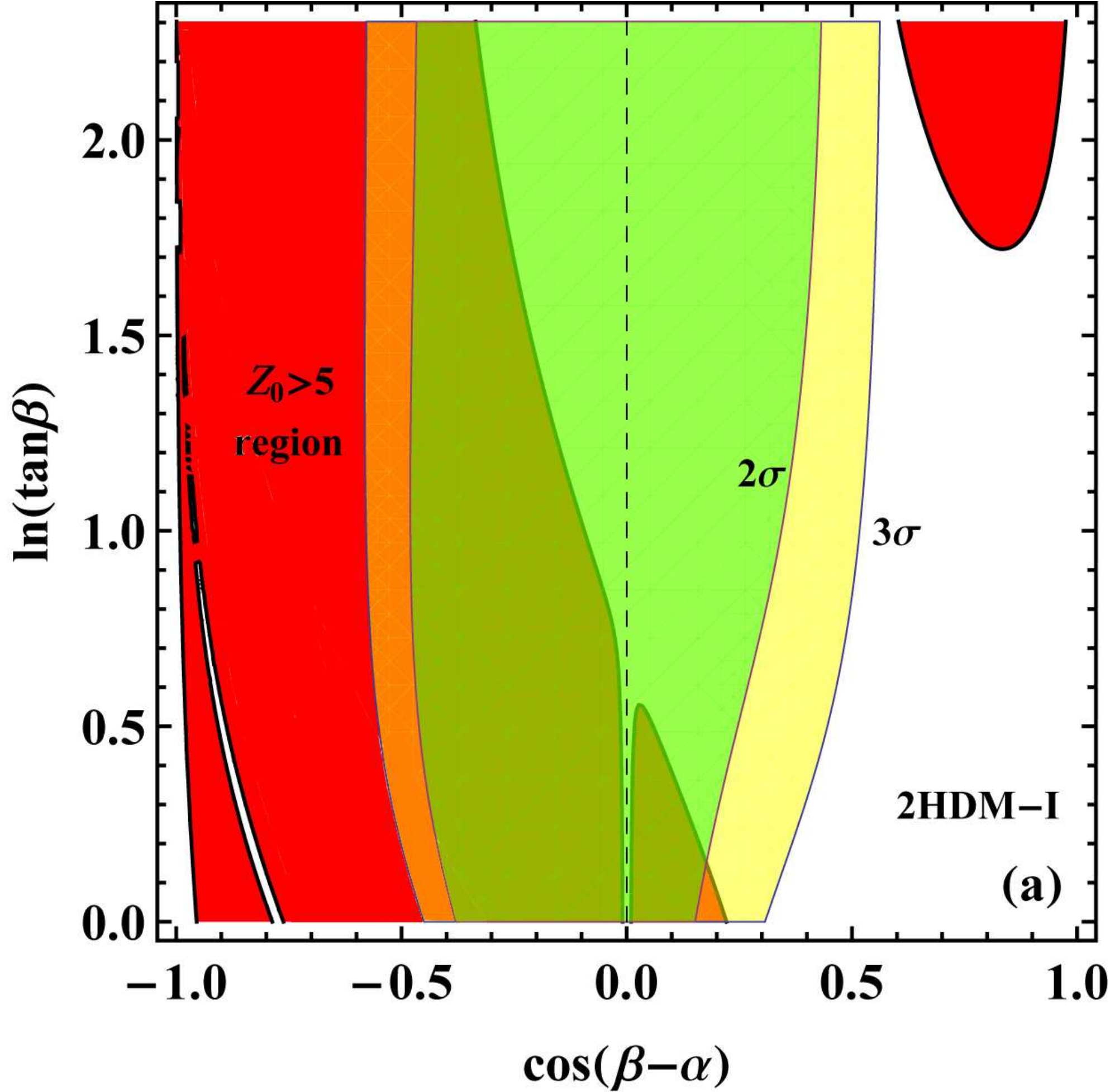}
\includegraphics[width=8.0cm,height=6.4cm]{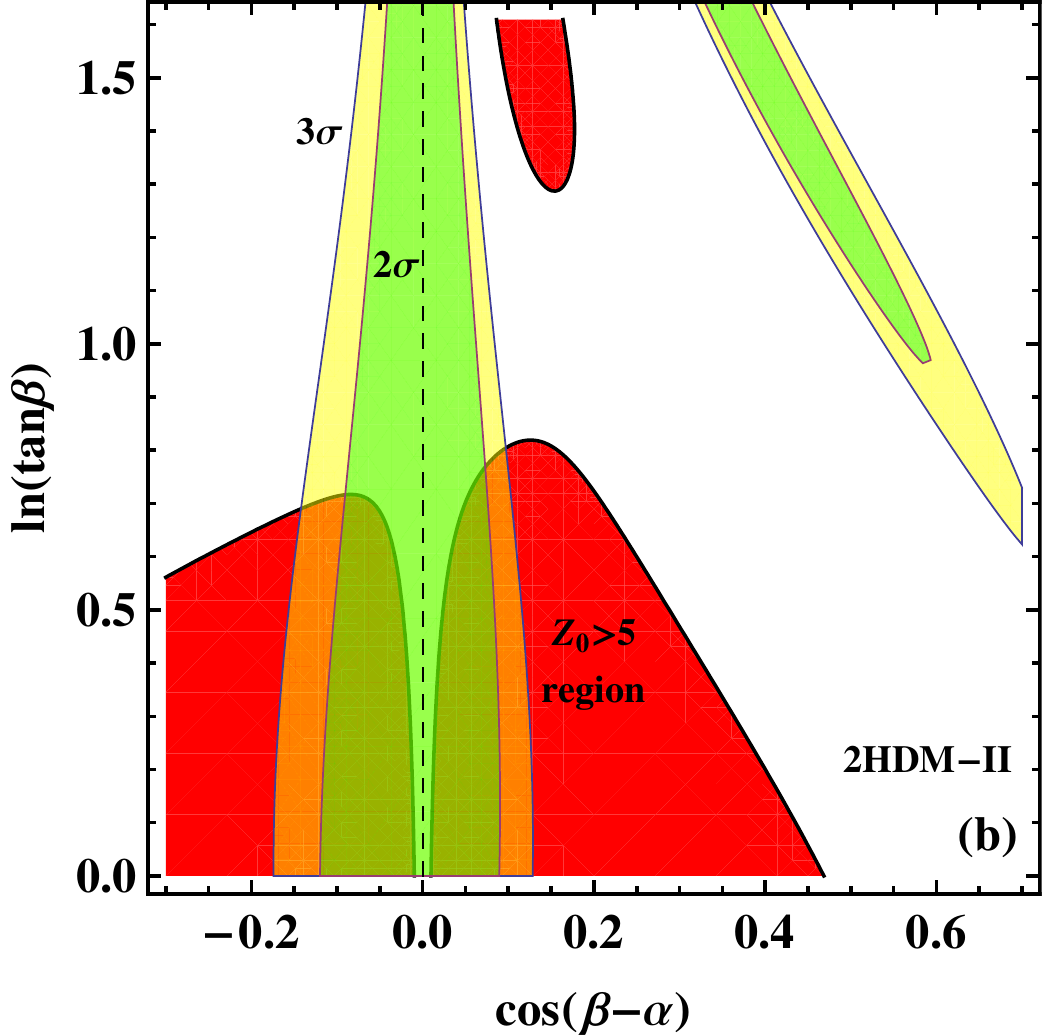}
\\[3mm]
\includegraphics[width=8.0cm,height=6.4cm]{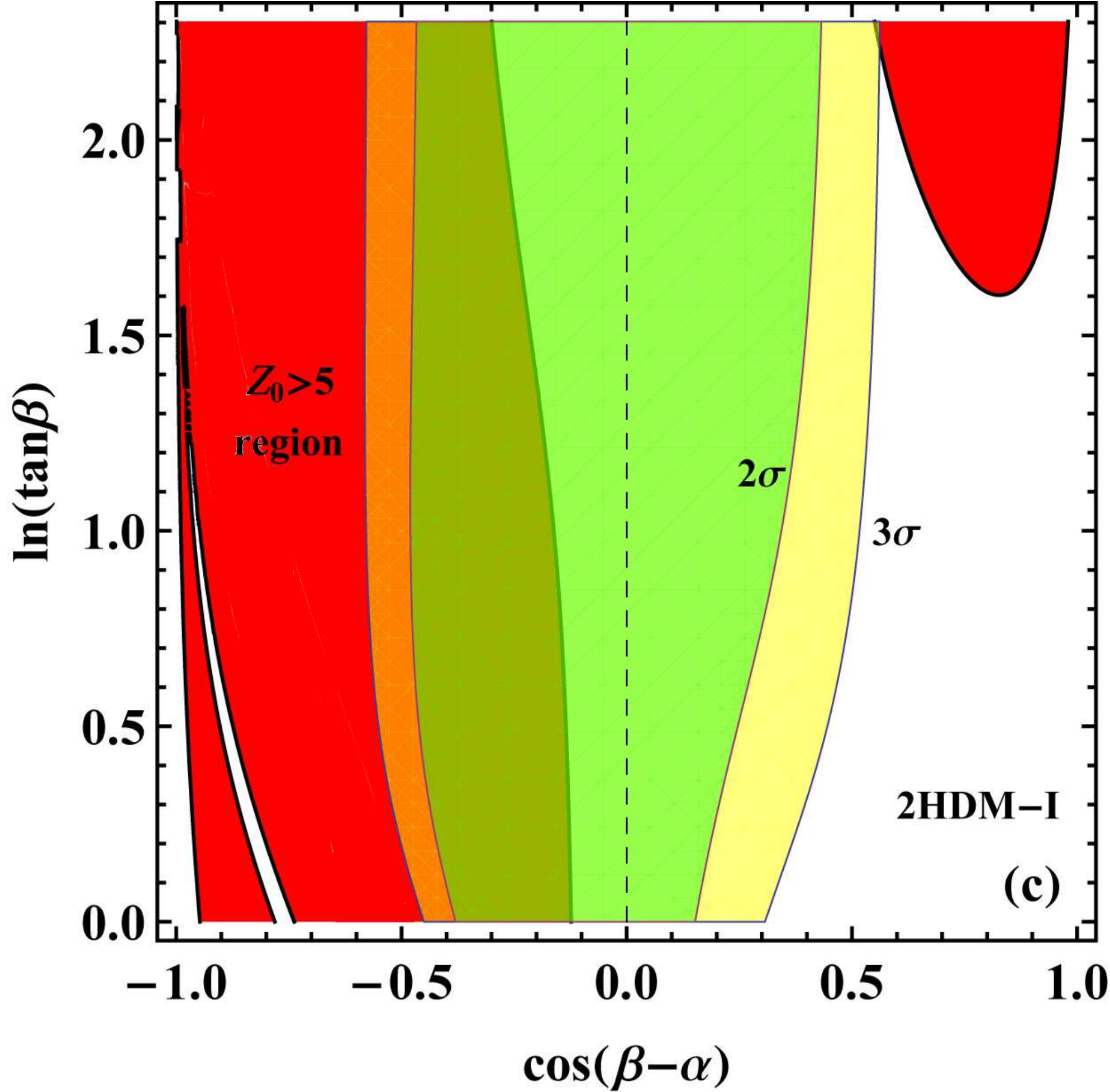}
\includegraphics[width=8.0cm,height=6.4cm]{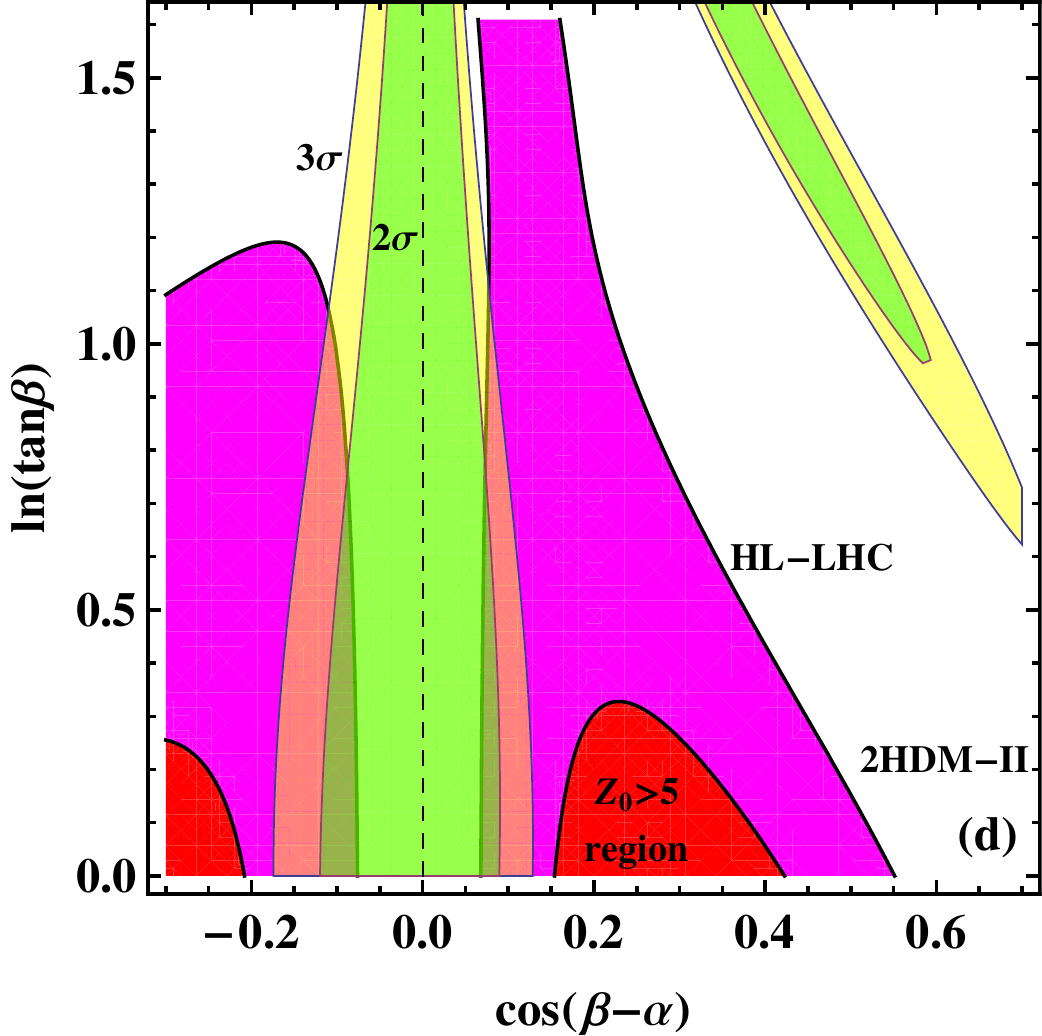}
\caption{LHC probe of 2HDM parameter space in $\,\cosba -\tanb$\, plane.
We impose projected sensitivity of the LHC Run-2 by requiring significance$(Z_0)> 5$\,
for the process $\,pp\!\to\!H\!\to\! hh\!\to\! WW^*\gamma\gamma\,$,\,
with an integrated luminosity ${\cal L}=300\,\text{fb}^{-1}$ at $\sqrt{s}=14$\,TeV.
All red contours correspond to significance$(Z_0)=5$\,
with ${\cal L}=300\,\text{fb}^{-1}$.\, Plots (a)-(b) [plots (c)-(d)] present the results for
$\,M_H^{}=300$\,GeV [$M_H^{}=400$\,GeV],
while plots (a) and (c) [plots (b) and (d)] give the results for 2HDM-I [2HDM-II].
In plot-(d), the pink contours ($Z_0=5$) show
a better probe with ${\cal L}=3\,\text{ab}^{-1}$ at the HL-LHC.
The green (yellow) contours present the $2\sigma$ ($3\sigma$) constraints from the
Higgs global fit of 2HDM-I [(a) and (c)] and 2HDM-II [(b) and (d)] at the LHC Run-1.
In all plots, we have sample inputs $\,(M_A^{},\,M_{\!12}^2)=\(500\GeV,\,-(180\GeV)^2\)\,$,\,
and the vertical dashed line denotes the alignment limit of the 2HDM.
}
\label{fig:9}
\end{centering}
\end{figure}

\vspace*{1mm}

We combine the significance($Z_0$) from
both pure leptonic channel $\,WW^*\gaga\to\ell\bar{\nu}\bar{\ell}\nu\gamma\gamma\,$
and semi-leptonic channel $\,WW^*\gaga\to q\bar{q}'\!\ell\nu\gamma\gamma$\,
at the LHC Run-2 with 300\,fb$^{-1}$ integrated luminosity.
For this analysis, the relevant mass-parameters of the 2HDM are
$\,(M_H^{},\,M_A^{},\,M_{\!12}^{})\,$.\, For demonstration, we will take the sample inputs,
$M_H^{}=300,400$GeV and $\,(M_A^{},\,M_{\!12}^2)=(500\GeV,\,-(180\GeV)^2)\,$.\,
With these, we have two remaining parameters in the 2HDM:
the mixing angle $\,\alpha\,$ and the VEV ratio $\,\tanb\,$.\,
In Fig.\,\ref{fig:9}, we impose projected sensitivity of
the LHC Run-2 by requiring significance$(Z_0)> 5$\,.\,
From this, we derive the red contours in the parameter space of $\cosba\!-\!\tanb$\, plane,
for 2HDM-I [plots (a) and (c)] and for 2HDM-II [plots (b) and (d)].
The plots (a)-(b) correspond to $\,M_H^{}=300$\,GeV\, and
plots (c)-(d) correspond to $\,M_H^{}=400$\,GeV\,.\,
This means that the LHC Run-2 with an integrated luminosity ${\cal L}=300\,\text{fb}^{-1}$
can probe the red contour regions in each plot of Fig.\,\ref{fig:9} with
a significance$(Z_0^{})> 5$\,.\,
It gives a discovery of the heavier Higgs boson $H$ (with 300\,GeV or 400\,GeV mass)
in the red regions of the 2HDM parameter space.

\vspace*{1mm}

In Fig.\,\ref{fig:9}, we further present the global fit for the lighter Higgs $h$\,(125GeV)
by using existing ATLAS and CMS Run-1 data, where the $2\,\sigma$ and $3\,\sigma$ contours of the
allowed parameter space are shown by the green and yellow shaded regions, respectively.
As we checked, our LHC global fit of the 2HDM is consistent with those in the literature
\cite{Djouadi:2013vqa}.
From this fit, we see that the parameter space favored by the current global fit
is around the alignment limit of 2HDM with $\,|\cosba |\lesssim 0.55\,$
for 2HDM-I and  $\,|\cosba |\lesssim 0.15\,$ for 2HDM-II.
But, 2HDM-II still has an extra relatively narrow
parameter region starting from  $\,\tanb\gtrsim 2$\,.\,

\vspace*{1mm}

Fig.\,\ref{fig:9}(a) has input $\,M_H^{}=300$\,GeV for 2HDM-I.
In this plot, the $\,Z_0^{}>5$\, region overlaps a large portion
of the parameter space favored by the current LHC global fit.
But, in Fig.\,\ref{fig:9}(b) for 2HDM-II,
the situation is different because the overlap becomes smaller
in the region $\,\cosba < 0\,$,\, and gets enlarged for $\,\cosba > 0\,$.\,
For the case of $\,M_H^{}=400$\,GeV in Fig.\,\ref{fig:9}(c),
the probed parameter space of 2HDM-I has sizable reduction, especially for
the region of $\,\cosba \gtrsim -0.05\,$,\,
in comparison with Fig.\,\ref{fig:9}(a) of $\,M_H^{}=300$\,GeV\,.\,
This is because the signal rate
decreases as $\,H\,$ becomes heavier [cf.\ Fig.\,\ref{fig:4}(a)].
On the other hand, for 2HDM-II, Fig.\,\ref{fig:9}(d) shows that
the $\,Z_0^{}>5$\, contours significantly shrink for $\,M_H^{}=400$\,GeV.\,
This is because the signal rate for 2HDM-II drops more rapidly
as Higgs mass rises to $\,M_H^{}=400$\,GeV in the small $\,\tanb$\, region
[cf.\ Fig.\,\ref{fig:4}(b)].
In this case, we see that the LHC Run-2
with $\,300\,\text{fb}^{-1}$ integrated luminosity
has rather weak sensitivities to the parameter space (shown by red contours),
and the red contours no longer overlap with the favored region
by the current LHC global fit (yellow and green contours).
We further analyze the probe from the upcoming High Luminosity LHC (HL-LHC)
with $\,3\,\text{ab}^{-1}$ integrated luminosity.
We find that the HL-LHC can significantly
extend the discovery reach of the parameter space of 2HDM-II,
as demonstrated by the pink contour regions ($Z_0^{}>5$) of Fig.\,\ref{fig:9}(d).

\vspace*{2mm}
\section{Conclusion}
\label{sec:5}
\vspace*{1.5mm}

After the LHC discovery of a light Higgs boson $h^0$(125GeV) at Run-1,
searching for new heavier Higgs state(s) has become a pressing task of the LHC Run-2.
Such heavier Higgs state(s) exists in all extended Higgs sectors and
can unambiguously point to new physics beyond the standard model (SM).

\vspace*{1.5mm}

In this work, we systematically studied the heavier Higgs boson $\,H^0\,$ production
with the new decay channel,
$\,pp\!\to\! H \!\to\! hh\!\to\! WW^*\gamma\gamma\,$,\, at the LHC Run-2.
In section\,\ref{sec:2}, we first analyzed the parameter space of the 2HDM type-I and type-II,
including the $Hhh$ cubic Higgs coupling (Fig.\,\ref{fig:1}).
We computed the decay branching fractions
and production cross section of the heavier Higgs boson $\,H\,$
at the LHC Run-2 over mass range $\,M_H^{}=250-600\,$GeV,\,
as shown in Fig.\,\ref{fig:2}--\ref{fig:4}.
Then, in section\,\ref{sec:3},
we analyzed both pure leptonic mode $\,WW^*\!\to\ell\bar{\nu}\,\bar\ell\nu\,$ and
semi-leptonic mode $\,WW^*\to q\bar{q}'\ell\nu\,$.\,
This channel has much cleaner backgrounds than the other process
$\,pp\!\to\! H\!\to\! hh\!\to\! \bb\gaga\,$.\,
We computed signal and background events using MadGraph5(MadEvent).
We applied {Pythia} to simulate hadronization of partons and
adopted {Delphes} for detector simulations. We followed the
ATLAS procedure for event selections and built kinematical cuts to efficiently
suppress the SM backgrounds. We analyzed various kinematical distributions
for pure leptonic and semi-leptonic decay channels in Fig.\,\ref{fig:6} and
Figs.\,\ref{fig:7}--\ref{fig:8} for three sample inputs of Higgs mass
$\,M_H^{}=(300,\,400,\,600)\,$GeV, respectively.
In Table\,\ref{tab:2}--\ref{tab:4},
we presented the signal and background rates of both channels under the kinematical cuts.
In section\,\ref{sec:4},
we combined the significance of pure leptonic and semi-leptonic channels, and
analyzed the LHC Run-2 discovery reach of $\,H\,$ as a probe of the
parameter space in 2HDM-I and 2HDM-II (Fig.\,\ref{fig:9}).
For comparison, we further presented the current Higgs global fit of the LHC Run-1 data
in the same plots. Finally, we note that it is hard to detect $H$
with mass above 600\,GeV at the LHC\,(14TeV) runs via di-Higgs production channel.
We find it valuable to extend our present LHC study to
the future high energy circular colliders $pp$($50\!-\!100$TeV) \cite{SPPC-FCC},
which are expected to further probe the heavier Higgs boson $H$ with mass up to
$O(1\!-\!5)$\,TeV range via $\,pp\to H\to hh\,$ production channel.

\vspace*{7mm}
\noindent
{\bf Acknowledgments}
\\[1.5mm]
We thank Weiming Yao for valuable discussions. We also thank
John Ellis, Yun Jiang, Tao Liu and Hao Zhang for useful discussions.
LCL and HJH are supported in part by National NSF of China
(under grants Nos.\,11275101 and 11135003) and
National Basic Research Program (under grant No.\,2010CB833000).
HJH acknowledges the support of visiting grants of IAS Princeton and Harvard University
during the finalization of this paper.
CD, YQF and HJZ are supported in part by Thousand Talents Program (under Grant No.\,Y25155AOU1).
This work is supported in part by the CAS Center for Excellence in Particle Physics (CCEPP).


\vspace{1mm}

\end{document}